\documentclass{pas}

\usepackage{multirow}
\usepackage{amsmath}
\usepackage{subcaption}
\usepackage{float}
\usepackage{booktabs}
\usepackage[table]{xcolor}
\usepackage{placeins}

\begin{document}

\lefttitle{OutThere Survey: Addressing $\mathrm{\xi_{ion}}$ and $\mathrm{f_{esc}}$ with a population of average galaxies at z$\sim$2.}
\righttitle{Ravi Jaiswar}

\jnlPage{1}{4}
\jnlDoiYr{2021}
\doival{10.1017/pasa.xxxx.xx}

\articletitt{Research Paper}

\title{OutThere Survey: Addressing $\mathrm{\xi_{ion}}$ and $\mathrm{f_{esc}}$ with a population of average galaxies at z$\sim$2.}


\author{Ravi Jaiswar $^{1,2}$}
\affil{$^1$International Centre for Radio Astronomy Research, Curtin University, Bentley WA, Australia}
\affil{$^2$ARC Centre of Excellence for All Sky Astrophysics in 3 Dimensions, Bentley WA, Australia}

\author{Anshu Gupta $^{1,2}$}

\author{Elisabete da Cunha $^{2,3}$}
\affil{$^3$International Centre for Radio Astronomy Research, University of Western Australia, Crawley WA, Australia}

\author{Cathryn M. Trott $^{1,2}$}
\author{Andrew Battisti$^{2,3}$}

\author{Isabel Perez $^{4,5}$}
\affil{$^4$Universidad de Granada, Departamento de F´ısica Te´orica y del Cosmos, Campus
Fuentenueva, Edificio Mecenas, E-18071, Granada, Spain }
\affil{$^5$Instituto Carlos I de F´ısica Te´orica y Computacional, Facultad de Ciencias, E-18071
Granada, Spain}

\author{Karl Glazebrook$^{6,7}$}
\affil{$^6$Centre for Astrophysics and Supercomputing, Swinburne University of Technology, P.O. Box 218, Hawthorn, VIC 3122, Australia}
\affil{$^7$JWST Australian Data Centre (JADC), Swinburne Advanced Manufacturing and Design Centre (AMDC), John Street, Hawthorn, VIC 3122, Australia}

\author{Lalitwadee Kawinwanichakij$^{6,7}$}

\author{Zackary L. Hutchens$^{8,9}$}
\affil{$^8$Department of Physics and Astronomy Elon University, 100 Campus Drive Elon, NC 27244, USA}
\affil{$^{9}$Department of Physics \& Astronomy, University of North Carolina Asheville, Asheville, NC 28804, USA}

\author{Ivelina Momcheva$^{10}$}
\affil{$^10$Max Planck Institute of Astronomy, Königstuhl 17, 69117 Heidelberg, Germany}

\author{Claudia Del.P Lagos$^{3,11}$}
\affil{$^{11}$Cosmic Dawn Center (DAWN), Denmark.}

\author{Themiya Nanayakkara$^{6,7}$}

\author{Colin Jacobs$^6$}

\author{Danilo Marchesini$^{12}$}
\affil{$^{12}$Department of Physics and Astronomy, Tufts University, 574 Boston Avenue, Medford, MA 02155, USA}

\author{David A. Wake$^{9,13}$}
\affil{$^{13}$Department of Physics and Astronomy, University College London, Gower Street, London WC1E 6BT, UK}

\author{Alice Shapley$^{14}$}
\affil{$^{14}$Department of Physics \& Astronomy, University of California, Los Angeles, 430 Portola Plaza, Los Angeles, CA 90095, USA}

\author{Gabriel Brammer$^{15,16}$}
\affil{$^{15}$Cosmic Dawn Center (DAWN), Denmark}
\affil{$^{16}$Niels Bohr Institute, University of Copenhagen, Jagtvej 128, DK-2200 Copenhagen N, Denmark}

\author{Raphael Hviding$^{10}$}

\author{Casey Papovich$^{17,18}$}
\affil{$^{17}$Department of Physics and Astronomy, Texas A\&M University, College Station, TX 77843-4242, USA}
\affil{$^{18}$George P. and Cynthia Woods Mitchell Institute for Fundamental Physics and Astronomy, Texas A\&M University, College Station, TX 77845-4242, USA}

\author{Ryan L. Sanders$^{19}$}
\affil{$^{19}$Department of Physics and Astronomy, University of Kentucky, 505 Rose Street, Lexington, KY 40506, USA; ryan.sanders@uky.edu}

\author{Rhea-Silvia Remus$^{20}$}
\affil{$^{20}$Universitäts-Sternwarte München, Fakultät für Physik, Ludwig-Maximilians Universität, Scheinerstr. 1, D-81679 München, Germany}

\author{Mengtao Tang$^{21,22}$}
\affil{$^{21}$Tsung-Dao Lee Institute, Shanghai Jiao Tong University, 1 Lisuo Road, Shanghai 201210, People’s Republic of China}
\affil{$^{22}$School of Physics and Astronomy, Shanghai Jiao Tong University, 800 Dongchuan Road, Shanghai 200240, People’s Republic of China}

\author{Daniel Stark$^{23}$}
\affil{$^{23}$Department of Astronomy, University of California, Berkeley, CA, United States}

\author{Vincente Estrada-Carpenter$^{24,25,26}$}
\affil{$^{24}$School of Earth and Space Exploration, Arizona State University, Tempe, AZ 85287, USA}
\affil{$^{25}$Beus Center for Cosmic Foundations, Arizona State University, Tempe, AZ 85287, USA}
\affil{$^{26}$Institute for Computational Astrophysics and Department of Astronomy \& Physics, Saint Mary’s University, 923 Robie Street, Halifax, NS B3H 3C3, Canada}

\author{Joshua Speagle$^{27,28,29,30}$}
\affil{$^{27}$Department of Statistical Sciences, University of Toronto, 9th Floor, Ontario Power Building, 700 University Avenue, Toronto, ON M5G 1Z5, Canada}
\affil{$^{28}$David A. Dunlap Department of Astronomy \& Astrophysics, University of Toronto, 50 St. George Street, Toronto, ON M5S 3H4, Canada}
\affil{$^{29}$Dunlap Institute for Astronomy \& Astrophysics, University of Toronto, 50 St. George Street, Toronto, ON M5S 3H4, Canada}
\affil{$^{30}$Data Sciences Institute, University of Toronto, 17th Floor, Ontario Power Building, 700 University Avenue, Toronto, ON M5G 1Z5, Canada}

\author{Kartheik Iyer$^{31}$}
\affil{$^{31}$Columbia Astrophysics Laboratory, Columbia University, 550 West 120th Street, New York, NY 10027, USA}

\author{ Jasleen Matharu$^{10,16,32}$}
\affil{$^{32}$Cosmic Dawn Center, Copenhagen, Denmark}





\history{(Received xx xx xxxx; revised xx xx xxxx; accepted xx xx xxxx)}

\begin{abstract}
Constraining the major contributors to the ionisation of the early universe is an ongoing endeavour of high-redshift galaxy research. We measure the ionising photon production efficiency and Lyman Continuum escape fraction for a sample of 230 intermediate redshift ($1.3<z<2.6$) sources observed as a part of the \textit{OutThere survey}; a pure-parallel, wide-area JWST/NIRISS survey with accompanying JWST/NIRCam, NIRISS and HST photometry. The low threshold emission selection criteria for this sample makes for a large and robust control, against which other works may be contrasted, particularly for low-mass galaxies above z$>$5. This control sample allows us to verify the correlations between ionising and spectral/physical properties suggested by previous studies. We find no significant correlations between the ionising photon production efficiency ($\mathrm{\xi_{ion}}$) with the UV slope, $\mathrm{M_{UV}}$, M$_*$ or sSFR. We do find that $\mathrm{\xi_{ion}}$ correlates with [OIII]5007\AA\, equivalent width (EW) (Spearman coefficient $\rho$ =0.24; p$< 4\times10^{-4}$) and H$\alpha$ EW ($\rho$ =0.63; p$<< 1\times10^{-6}$) hold even at low EW albeit with more scatter. We also find that our novel approach to determining the ionising photon escape fraction $\mathrm{f_{esc}}$ results in values within theoretical ranges (0-10\%) though vary substantially in comparison to the empirical results (median $\mathrm{f_{esc}} = 0.9\%^{+1.1}_{-0.5}$ including non-detections, median $\mathrm{f_{esc}} = 1.9\%^{+8.9}_{-1.8}$ above a $0.01\%$ threshold). We find that this escape fraction method has consistently significant correlations with the redshift, SFR and M$_{UV}$ and sample-dependent correlations with [OIII]5007\AA\,EW,H$\alpha$ EW and stellar mass.  
\end{abstract}

\begin{keywords}
Galaxy Evolution, Epoch of Reionisation, Lyman Continuum, Ionising photon production
\end{keywords}

\maketitle

\section{Introduction} \label{sec:intro}

The Epoch of Reionisation (EoR) marks the period during which the intergalactic medium transitioned from predominantly neutral to ionised. Constraining the timescale and drivers of this process requires characterising the sources of ionising photons to constrain their possible contributions, even though direct detection of Lyman‑continuum (LyC) emission is prevented by the opacity of the early intergalactic medium (IGM) \citep{Robertson2010,Robertson2013}. It is yet unclear if the major contributors to reionisation were massive bright galaxies \citep{Sharma2018, Naidu2020} or more numerous faint  ones (M$_{UV}>-18$) \citep{Ciardi2003,Bouwens2012, Robertson2010}.

Since the launch of JWST, the ionising photon production efficiency parameter, $\mathrm{\xi_{ion}}$, has become a central tool for quantifying the LyC contribution of galaxies to reionisation. It has been measured for direct EoR sources \citep{Simmonds2023, Boyett2024, Pahl2024,Harshan2024,Papovich2026}, for EoR analogues at intermediate \citep{Tang2019, Jaiswar2024} and low \citep{Cardamone2009, Izotov2018} redshifts and for specially selected sources \citep{Stark2014,Gupta2023}. 

Low‑mass or ``faint" galaxies remain compelling candidates for driving reionisation due to their early formation, high number densities (thus large UV luminosity density $\rho_{UV}$) \citep{Balu2023,Cook2024}, elevated specific SFRs at high redshift (z$>$6) \citep{Robertson2023} in comparison to the low redshift equivalents, and the potential for shallower gravitational wells that promote LyC escape via stellar wind and radiation driven feedback \citep{Izotov2018, Finkelstein2019}. The Ly$\alpha$ forest exhibited by quasars in the EoR has always suggested a significant density of neutral Hydrogen (HI) gas toward the recent end of the EoR, which is better modelled by low-mass contributors \citep{Kulkarni2019}.  

However, the tendency to select interesting sources to analyse can lead to biases in our understanding of ionising emission properties. Galaxies exhibiting bright emission lines \citep{Bunker2023-gnz11,Carniani2024,Helton2025}, lacking dust obscuration \citep{Jaiswar2024,Setton_2025}, having atypical continuum profiles \citep{Cameron2023}, or those interacting with other galaxies \citep{Gupta2023,Mascia2025} may not represent the general population but are over-represented in this field of study. Such samples are highly informative, but may not represent the underlying population, and analyses that rely exclusively on them risk biasing the inferred ionising budgets and obscuring the true evolution of ionising parameters \citep{Simmonds2024}.

\begin{figure*}
    \centering
    \includegraphics[width=0.95\textwidth,height=0.95\textheight,keepaspectratio]{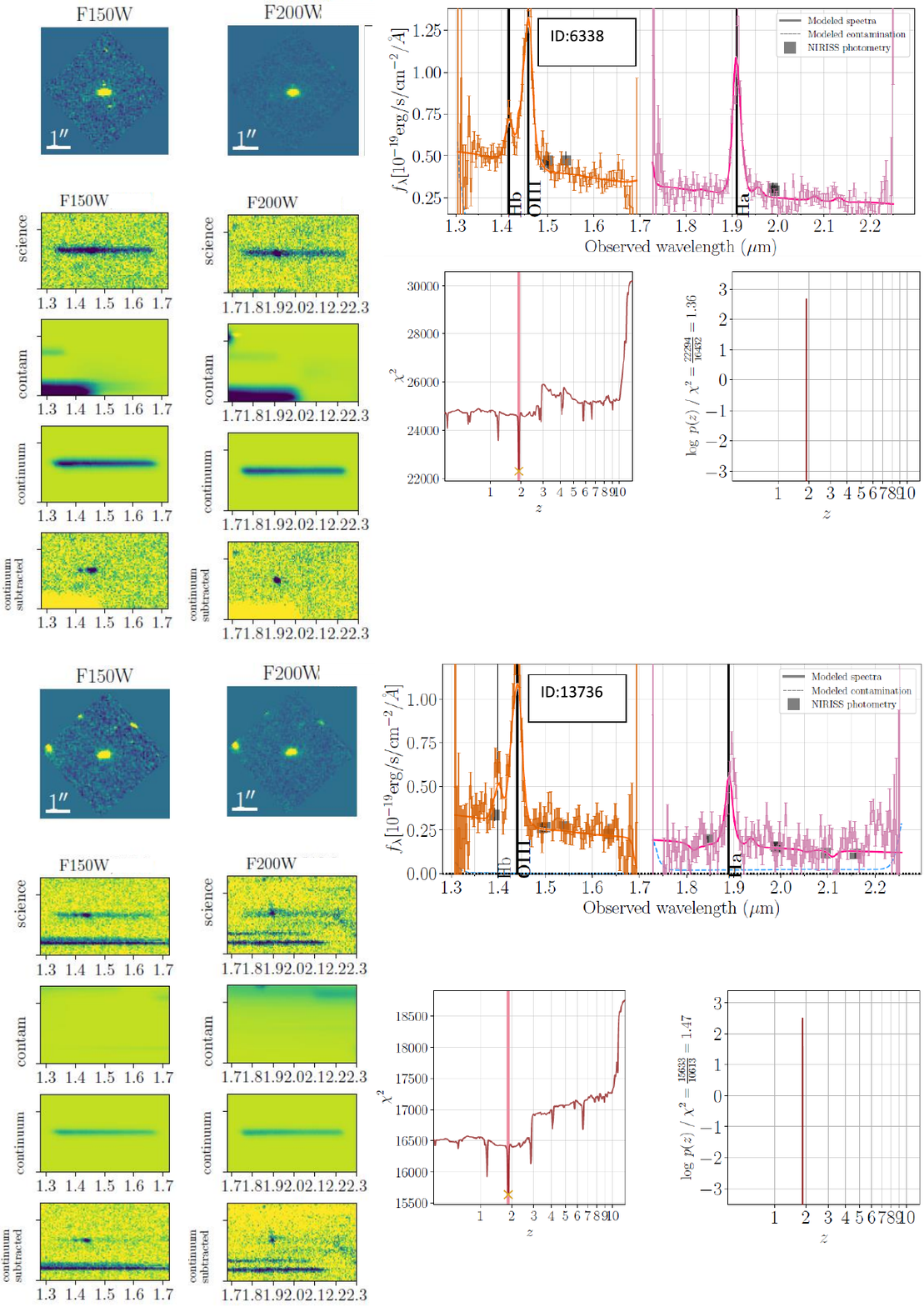}
    \caption{Sample 1D and 2D spectra in both F150W and F200W and redshift likelihood fit for galaxy 6338 (top) and 13736 (bottom), with highly certain z fits and well isolated emission lines. 6338 is given as an ideal example of the Grizli reduction pipeline while 13736 is given as a reasonably well fit example where the remaining contamination does not overlap spatially with our source. These are included in the final sample.}
    \label{fig:spectra_good}
\end{figure*}

\begin{figure*}
    \centering
    \includegraphics[width=0.95\textwidth,height=0.95\textheight,keepaspectratio]{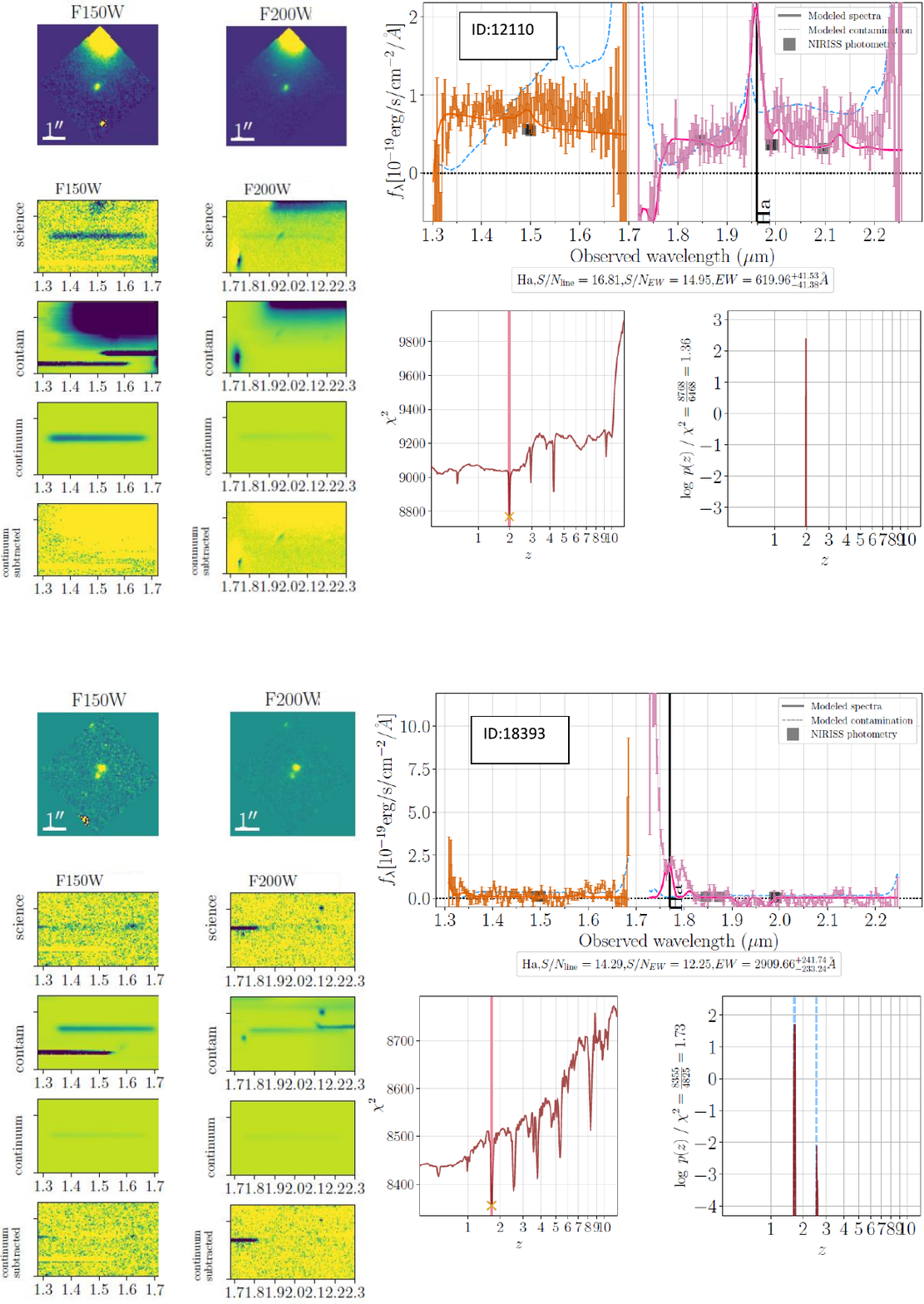}
    \caption{Sample 1D and 2D spectra in both F150W and F200W and redshift likelihood fit for galaxy 12110 (top) and 18393 (bottom). 12110 is given as a borderline example of the of the Grizli reduction pipeline while 18393 is given as a poorly constrained example. 12110 is considered borderline as while the z fit is certain it appears to derive the H$\alpha$ emission line from a feature that is present in the contamination model. 18393 is considered a failed fit as the H$\alpha$ line falls on the filter edge and has two major z fit peaks. These are not included in the final sample.}
    \label{fig:spectra_bad}
\end{figure*}

Large direct studies of faint, high-redshift (M$_{UV}>-18$, z$>6$) sources with good constraints are scarce \citep{Harshan2024} and creating a true analogue sample at lower redshifts without a proper calibration sample is difficult because of this. However, constraining the relationship of probes for the escape fraction of LyC photons ($f_{esc}$) and $\mathrm{\xi_{ion}}$ with these parameters is becoming more accessible with the depth of new programs.   
\begin{figure}
    \centering
    \begin{subfigure}{\linewidth}
        \centering
        \includegraphics[width=\linewidth]{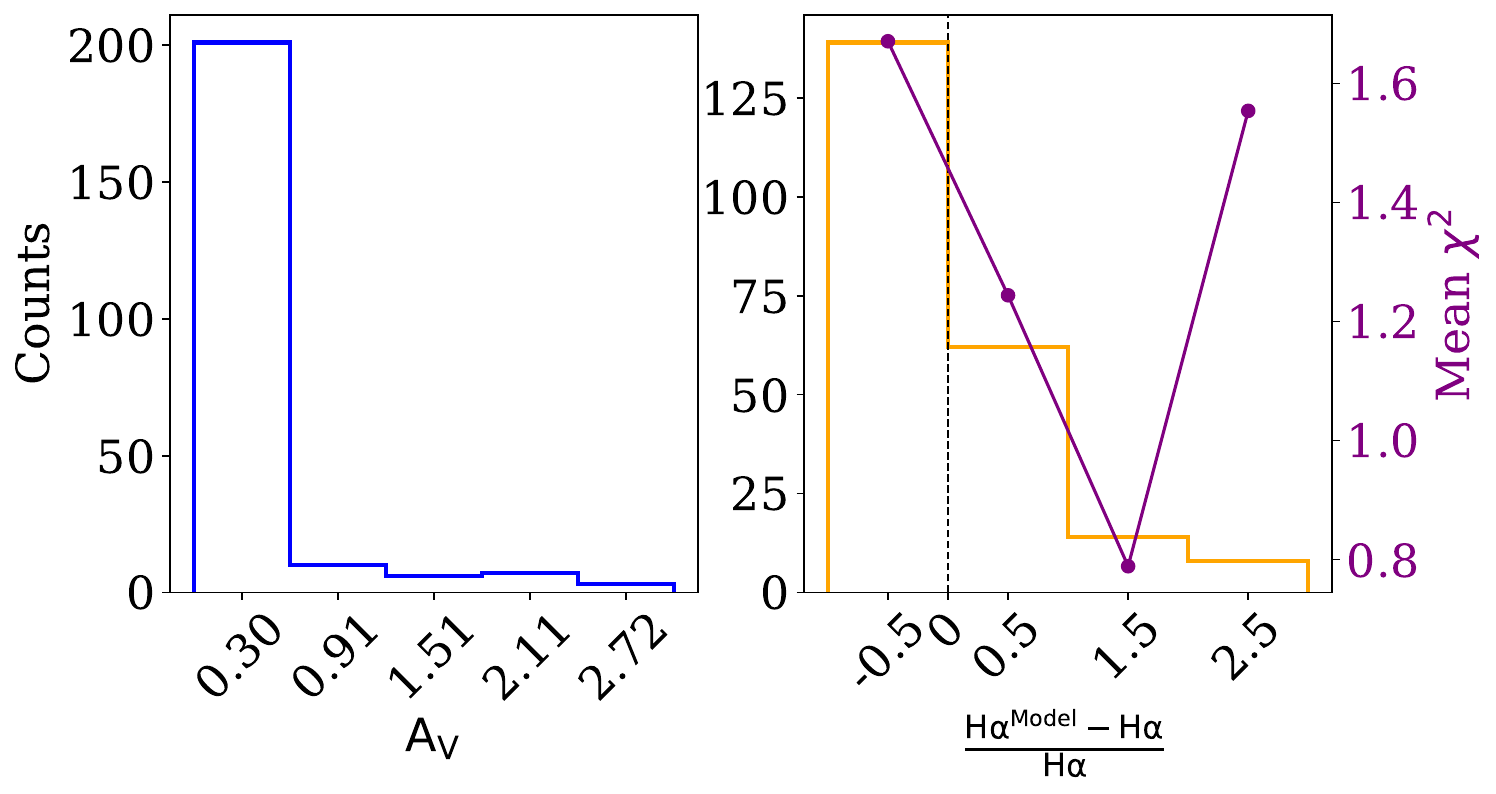}
        \caption{}
        \label{fig:quality_a}
    \end{subfigure}

    \vspace{0.8em}

    \begin{subfigure}{\linewidth}
        \centering
        \includegraphics[width=\linewidth]{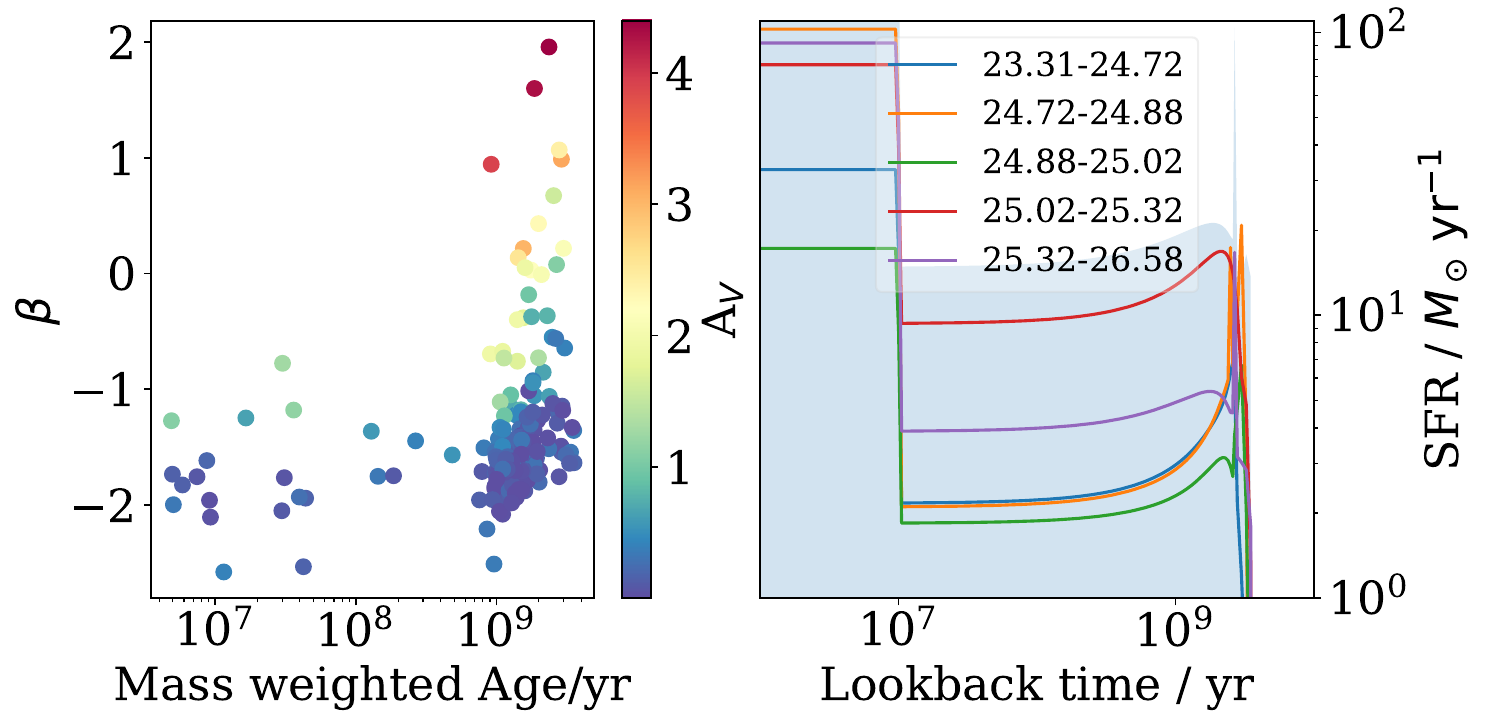}
        \caption{}
        \label{fig:quality_b}
    \end{subfigure}
    \caption{Quality control and diagnostic tests for our 230 OutThere galaxy sample. (a) Left) Dust attenuation per galaxy count determined using \texttt{BEAGLE} and the CF00 dust law. (a) Right) Relative difference of the modelled $H\alpha$ emission line flux to the OutThere NIRISS $H\alpha$ flux (orange) over-plotted with the separation of the modelled SED to the photometry (purple, reduced $\chi^2$). (b) Left) UV slope correlation with stellar ages and dust attenuation. (b) Right) Star formation history averages over 5 bins of $\mathrm{\xi_{ion}^{HII}}$. We find that most of our \texttt{BEAGLE} SED galaxies experience low dust attenuation and underestimate the $H\alpha$ flux. The UV slope is more strongly related to the dust attenuation than to the stellar age. We find no clear starburst behaviour responsible for particularly high $\mathrm{\xi_{ion}^{HII}}$.}
    \label{fig:QC}
\end{figure}

\begin{figure}
    \centering
    \includegraphics[width=\linewidth]{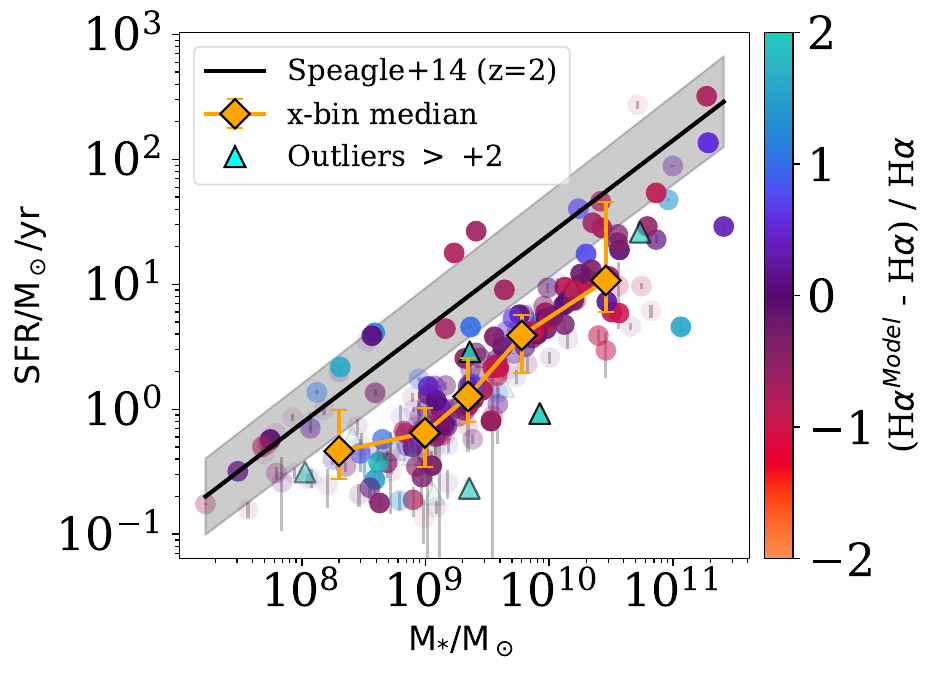}
    \caption{Stellar Mass vs Star Formation rate (averaged over 100 Myr) for our OutThere sample of 230 galaxies. Orange diamonds correspond to equal-population bin medians, and their error-bars reflect the 16-84th percentile ranges. The main sequence relation for z=2 from \cite{Speagle14} is indicated in black with upper and lower limits indicated in grey. Blue diamonds reflect galaxies where $\mathrm{\frac{H\alpha^{Model}-H\alpha}{H\alpha}> +2}$. Our sample lies largely below the Main Sequence.}
    \label{fig:MainSequence}
\end{figure}
\begin{figure}
    \centering
    \includegraphics[width=\linewidth]{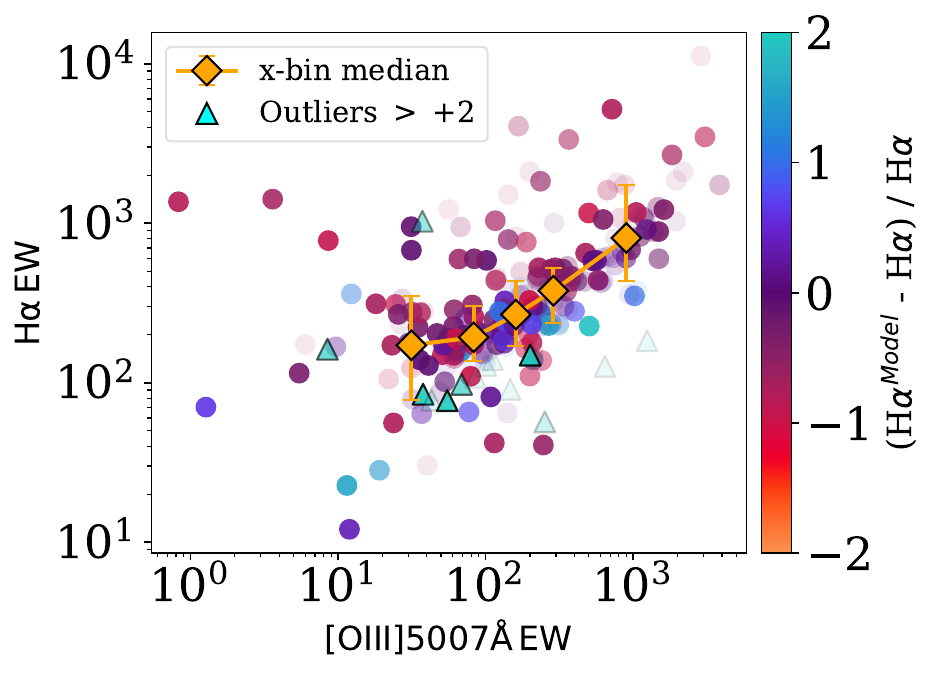}
    \caption{$\mathrm{[OIII]5007}$\AA\,EW vs $\mathrm{H\alpha}$\AA\,EW for our OutThere sample of 230 galaxies. Orange diamonds correspond to equal-population bin medians (excluding limits), and their error-bars reflect the 16-84th percentile ranges. Blue diamonds reflect galaxies where $\mathrm{\frac{H\alpha^{Model}-H\alpha}{H\alpha}> +2}$. Our sample spans a notable range in both EWs, with $\sim$23\% within the extreme emission line definition of \cite{Jaiswar2024}.}
    \label{fig:O3Ha}
\end{figure}

In this paper we present a new sample from the OutThere survey, a wide‑area pure‑parallel spectroscopic James Webb Space Telescope (JWST) program. Our selection identifies $z\sim2$ [OIII]5007 and H$\alpha$ emitters across the full range of detected equivalent widths, enabling us to probe the low‑EW and low‑ionising‑efficiency regimes that are typically inaccessible or unreported in similar high‑redshift studies. The aim of this work is to construct a large, low–entry‑barrier control sample at intermediate redshift (z$\sim2$) that can serve as a baseline for interpreting bright and faint emission‑line populations observed deep into the EoR. The optical coverage and high‑quality photometry available at these redshifts allow us to constrain stellar and ionising properties more reliably than is possible for $z>6$ systems, providing a reference against which EoR measurements can be placed in context. We note that, because spectroscopic redshifts require identifiable emission features, the sample necessarily favours galaxies with detected [OIII]5007 and H$\alpha$; however, within this constraint we include the full observed range of line strengths to minimise additional selection bias.

\begin{figure*}
    \centering
    \begin{subfigure}{\linewidth}
        \centering
        \includegraphics[width=0.95\linewidth]{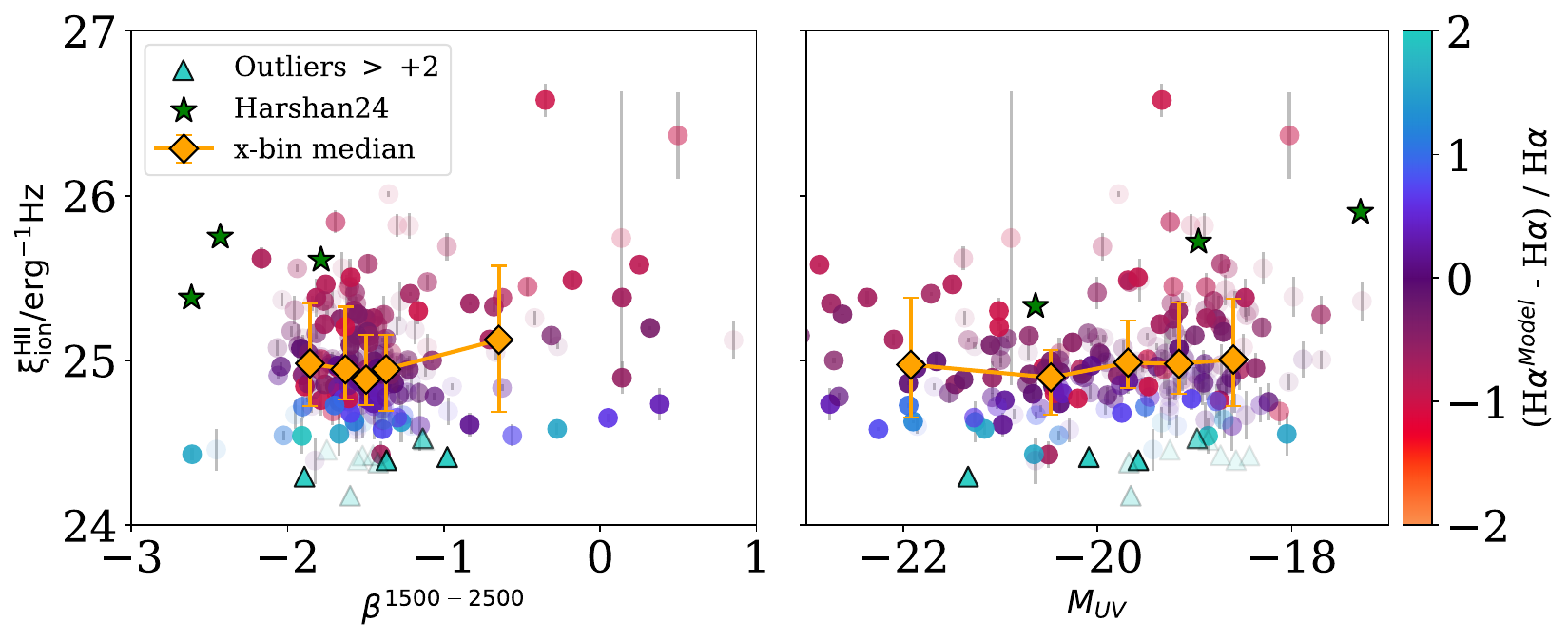}
        \caption{}
        \label{fig:paper3_a}
    \end{subfigure}

    \vspace{0.8em}

    \begin{subfigure}{\linewidth}
        \centering
        \includegraphics[width=0.95\linewidth]{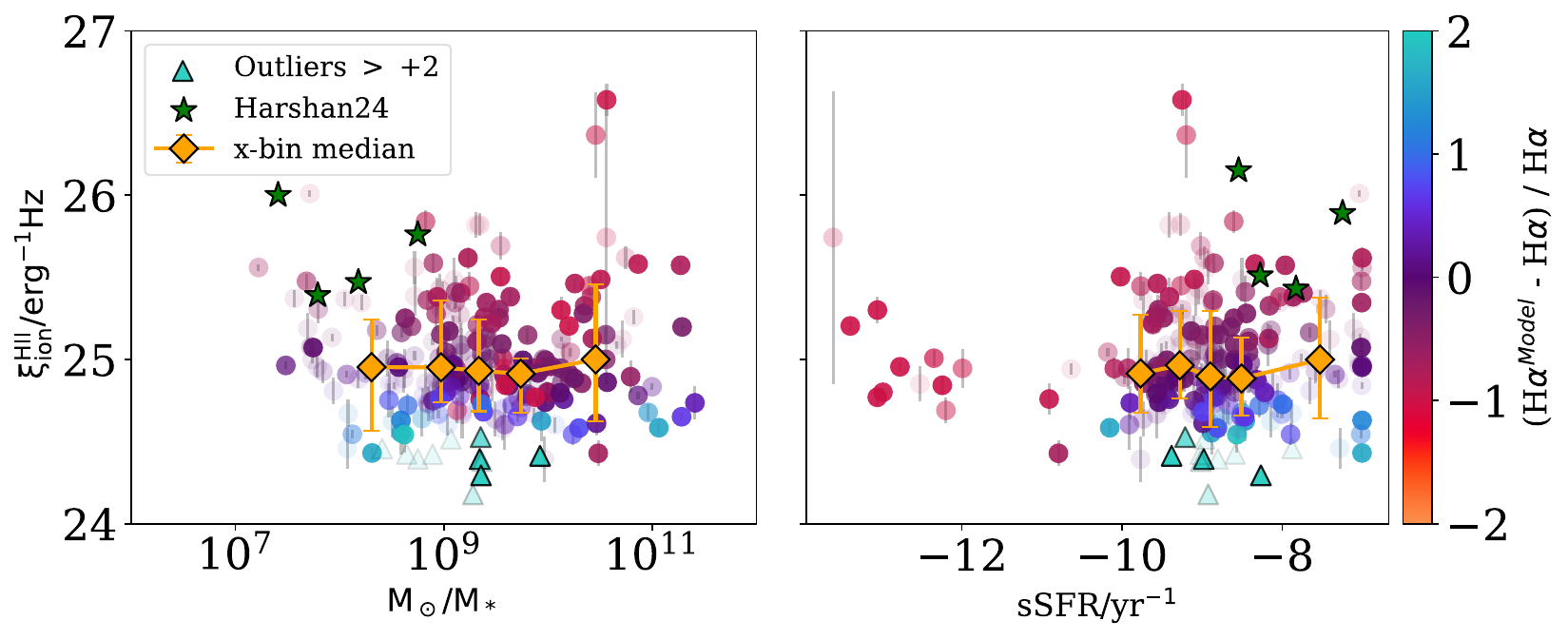}
        \caption{}
        \label{fig:paper3_b}
    \end{subfigure}
    \caption{ (a) Left) UV slope vs $\mathrm{\xi_{ion}^{HII}}$  (a) Right) M$_{UV}$ vs $\mathrm{\xi_{ion}^{HII}}$ (b) Left) Stellar mass M$_*$ vs $\mathrm{\xi_{ion}^{HII}}$ (b) Right) sSFR vs $\mathrm{\xi_{ion}^{HII}}$. Opacity of data-points inversely proportional to \texttt{BEAGLE} SED reduced $\chi^2$ so that better fits are more opaque. Orange diamonds correspond to equal-population bin medians, and their error-bars reflect the 16-84th percentile ranges. Points from \cite{Harshan2024} plotted as green stars. We find no significant Spearman correlations between $\mathrm{\xi_{ion}^{HII}}$ and any of these parameters.}
    \label{fig:xion-bmuv}
\end{figure*}

\begin{figure*}
    \centering
     


        \includegraphics[width=0.95\linewidth]{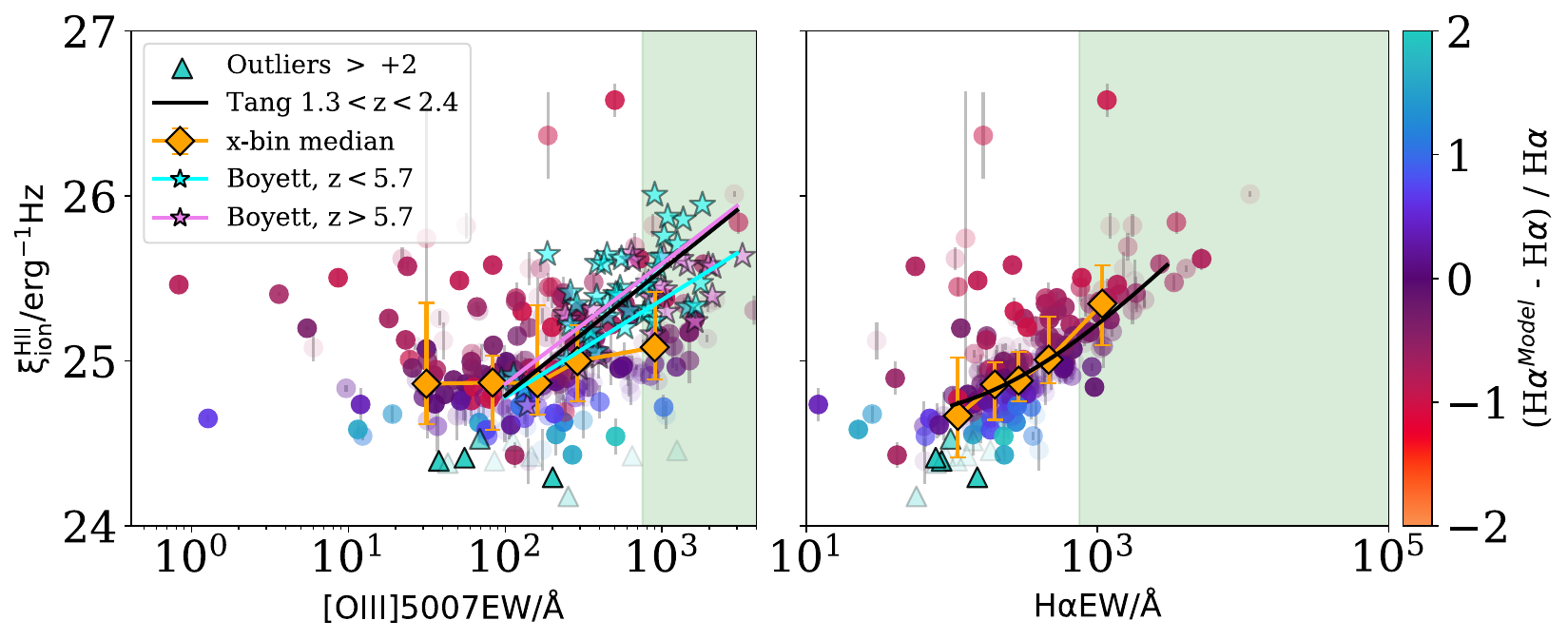}
        \label{fig:paper3_d}
    
    \caption{(Left) [OIII]5007 EW vs $\mathrm{\xi_{ion}^{HII}}$ and (Right) H$\alpha$ EW vs $\mathrm{\xi_{ion}^{HII}}$. Opacity of data-points inversely proportional to \texttt{BEAGLE} SED reduced $\chi^2$ so that better fits are more opaque. Orange diamonds correspond to equal-population bin medians, and their error-bars reflect the 16-84th percentile ranges. Region above EW=750\AA\, indicated by the green shaded region as in \cite{Boyett2024}, and points derived from \cite{Boyett2024} in pink and cyan stars. Trends from \cite{Tang2019,Boyett2024} indicated by black, pink and cyan lines. We find significant Spearman correlations between $\mathrm{\xi_{ion}^{HII}}$ and each of [OIII]5007 EW ($\rho$ = 0.26) and H$\alpha$ EW ($\rho$ = 0.65) in our samples.}
    \label{fig:o3}
\end{figure*}

\section{Data and Methods}

The rate of ionisation N$_{ion}$ at some given epoch is parameterised by the fraction of escaping Lyman Continuum light ($\mathrm{f_{esc}}$), the ionising photon production efficiency ($\mathrm{\xi_{ion}}$) and the number density of ionising sources within a co-moving volume, $\rho_{UV}$ \citep{Robertson2010}. Here we explain our derivation of the $\mathrm{f_{esc}}$ and $\mathrm{\xi_{ion}}$ parameters that are used in this work. We also explore the OutThere spectroscopic survey data and the combined photometric catalogues used to derive these parameters.

\subsection{OutThere Survey}
The OutThere survey is a JWST pure-parallel program using the Near Infrared Imager and Slitless Spectrograph (NIRISS) instrument grism spectroscopy. As of writing, this survey has collected data from 120/127 expected fields over 425 hours, making it the largest such JWST program to date. This work focuses on a selection of 12 pointings over 48 sq.arcmin  in the Hubble Ultra-Deep Field-North (HUDF-N) field. We use the F150W and F200W filters which give the reddest spectra up to 2.2$\mu$m with a depth of more than 12.5 hours.

\subsection{Photometry and selection}\label{subsec:OutThere}
To constrain the physical and selection parameters of these galaxies, we use the combined photometric data publicly available from the JADES \citep{jades}, PANORAMIC \citep{panoramic}, CONGRESS \citep{congress}and SAPPHIRES \citep{sapphires} programs in the HUDF-N field. This internal, multi-band catalogue was constructed using JWST NIRISS, Hubble Space Telescope (HST) ACS/WFC and JWST Near Infrared Camera (NIRCam) Grizli-reduced imaging from these programs spanning 24 filters from F435W to F460M.

Galaxies were selected to contain detected [OIII]5007 and H$\alpha$ emission lines between $1.6<z<2.5$ as well as at least 12/24 filters in order to consider only well sampled spectral energy distributions (SEDs) and so that the filters captured both lines. This provides an initial sample of 371 sources. Adding a SN$>3$ cut on H$\alpha$ reduces this sample to 277. We found 3 out of our 277 galaxies did not converge on a solution from SED fitting with the priors (see Table \ref{tab:Priors}) used by the full sample. To maintain consistent working priors across the entire sample, these were removed. A further 41 galaxies were eliminated due to poorly reduced spectra as detailed in section \ref{subsec:OutThere_spectroscopy} (see Figures \ref{fig:spectra_good} and \ref{fig:spectra_bad}), giving a sample of 233 galaxies. Figure \ref{fig:QC} shows some general properties (SFH, dust/age correlation) and diagnostics regarding the SED modelling. 

This photometry was fitted using the \texttt{BEAGLE} SED fitting code \citep{Chevallard2016,Gutkin2016}, a Bayesian tool which interprets the physical parameters of the galaxy from this fit. It contains the BC16 SPS model, which is an updated version of \cite{Bruzual2003}, CLOUDY photo-ionisation \citep{Ferland2017}, CF00 dust model \citep{Charlot2000}, delayed-tau star formation history and a free parameter constant SFH at 10 Myr. \texttt{BEAGLE} self consistently considers dust within the nebular emission model when using two component dust models (such as CF00) hence our choice of model. 

We determine the quality of these fits using the reduced $\chi^2$ and remove sources above $\chi^2$=3.5 or which had $\chi^2$=0 (3 sources). This sets the final sample as 230 galaxies. The fitting uses the prior ranges in Table \ref{tab:Priors}.

\begin{table}[]
    \centering
    \begin{tabular}{c|c|c}
         Description&Parameter&Range\\
         Star formation timescale&$\tau$/log(yr)& 7, 9.4\\   
         Metallicity&Z/log(Z$_\odot$)&-2.6, 0.6\\
         Stellar Mass&M/log(M$_\odot$)&6, 13\\
         Star formation rate&SFR/log(M$_\odot$/yr)&-3, 6\\
         Effective ionisation parameter &nebular\_logU&-4,-1\\
         Effective dust/Z mass ratio&nebular\_$\xi$&0.1, 0.5\\
         V-band attenuation optical depth&$\tau_V\mathrm{eff}$&0.001, 5\\
         optical depth ISM fraction&$\mu$&0.4\\\end{tabular}
    \caption{Prior ranges set for each parameter in the \texttt{BEAGLE} SED model. All ranges use uniform distributions.}
    \label{tab:Priors}
\end{table}

\subsection{OutThere Spectroscopy}\label{subsec:OutThere_spectroscopy}

At the time of publication the OutThere survey is limited to its internal data releases. Data processing and calibration are still in progress, particularly relating to the trace and wavelength calibrations. Spectral extraction and emission line measurements are affected by NIRISS contamination and should be interpreted with appropriate caution. Full descriptions of the reduction, calibration steps, and sensitivities will be presented in forthcoming OutThere papers once the pipeline is stabilised. As a result, we visually classify individual sources to eliminate failed emission line fits, bad redshift constraints and poorly modelled contamination. This process eliminated 41/274 sources. 

The visual classification was performed by grouping galaxies into `good' and `bad' examples of forward-modelled, slitless NIRISS data, where the borderline galaxies were grouped into either category. The visual selection criteria of good and borderline-good sources are exemplified in Figure \ref{fig:spectra_good} while Figure \ref{fig:spectra_bad} exemplifies borderline-bad and bad sources from top to bottom. Figure \ref{fig:spectra_good} (top) shows an ideal fit, where the Grizli pipeline has modelled both the contamination and the continuum before calculating the identified emission lines, and the redshift fit is of high certainty. The bottom panel shows an example of a source accepted despite contamination and residual continuum being clearly visible in the final 2D spectrum. As the contamination is spatially distinct in the 2D spectrum from the emission lines of interest, and as the redshift certainty appears significant, this galaxy was also included in the final sample. Figure \ref{fig:spectra_bad} (top) shows an example source with significant contamination in both F150W and F200W. While the model appears to eliminate this contamination and the underlying continuum, this source appears to derive its H$\alpha$ flux from an elongated feature that is present in its contamination model which has not been properly masked. To ensure the purity of the H$\alpha$ flux in our sample, this source was eliminated. The bottom panel shows a source with poorly modelled contamination that is not distinct from the galaxy and appears on the edge of the filter. Furthermore, it has multiple likely redshift solutions and so was rejected from the final sample. 

\subsection{Dust correcting H$\alpha$}
At the spectral resolution of our NIRISS spectra, H$\alpha$ (6563\AA) is blended with the double peaked [NII] (6548, 6583), which would overestimate its flux \citep{Faisst2018}. We utilise the photo-ionisation modelling of our SEDs to proportionally subtract these emission line fluxes from the H$\alpha$ flux as:

\begin{equation}
    \mathrm{F_{H\alpha-NII} = F_{H\alpha}^{spec} \times \left(1- \frac{F_{[NII]_{6548}^{Model}}+F_{[NII]_{6583}^{Model}}}{F_{{H\alpha}^{Model}}}\right)}
\end{equation}
We then correct the H$\alpha$ emission for dust absorption by considering the CF00 attenuation model using the \texttt{BEAGLE} SED derived A$_V$ value as:

\begin{equation}
    \begin{split}
        \mathrm{\tau_V = \frac{A_V}{1.086 \times 0.44}}, \\
        \mathrm{k_{H\alpha} = \left(\frac{6563}{5500}\right)^{-1.3}},\\
        \mathrm{H\alpha_{int}=H\alpha \times 10^{(0.4 \times \tau_V \times k_{H\alpha})}},
    \end{split}
\end{equation}
which uses the reference wavelength of 5500\AA\, and the birth cloud (BC) absorption coefficient of n=1.3 from the CF00 dust attenuation model. We use the factor of 0.44 to convert from stellar to nebular A$_V$ \citep{Calzetti2000,Battisti2016}. All reported H$\alpha$ refers to this dust corrected version. Figure \ref{fig:QC}'s top left panel shows the \texttt{BEAGLE} derived A$_V$ distribution for this sample (1$\sigma$ A$_V = 0.14^{+0.37}_{-0.11}$). It is worth noting the dependence of these results on the exact implementation of dust corrections, which were not explored in this work. Applying a stronger dust attenuation factor makes many of the noted relations correlate where they previously did not, as the most dusty galaxies are very dependent on this factor. An unorthodox method to correct for dust was explored using the \texttt{BEAGLE} model, such that the modelled emitted and observed H$\alpha$ correct the spectroscopic H$\alpha$ as: \begin{equation}
   \mathrm{ H\alpha^{corr} = H\alpha^{spec} \times \frac{H\alpha^{em}}{H\alpha^{obs}}},
\end{equation}
though ultimately this was deemed to introduce too large a model dependence. Alternatively, an argument could be made for the factor n=1.3 in the calculation of $\tau_{H\alpha}$ to instead be n=0.7 for the diffuse ISM attenuation following the birth cloud. Yet additionally, the factor of 0.44 used to convert the stellar A$_V$ to a nebular value is not well supported \citep{Reddy2015}. These uncertainties in the method should be noted and revisited with a deeper understanding of the underlying processes causing them.

\subsection{Determining $\mathrm{\xi_{ion}}$}\label{subsec:xion}
The ionising photon production efficiency quantifies the intrinsic production rate of hydrogen-ionising photons relative to the non-ionising UV continuum luminosity, which traces the young stellar population. The primary derivation method of $\mathrm{\xi_{ion}}$ using H$\alpha$ emission (hence $\mathrm{\xi_{ion}^{HII}}$) \citep{Simmonds2023} follows:
\begin{equation}
    \begin{split}
         N(H^0) = L(H\alpha)\times7.28\times10^{11}, \\
        \log_{10}(\xi_{ion}^{HII})=\log_{10}\left(\frac{N(H^0)}{L_{UV}}\right)
    \end{split}
\end{equation}
where N(H$^0$) is the total rate of hydrogen-ionising photons ($s^{-1}$) produced, L(H$\alpha$) is the H$\alpha$ luminosity (in erg s$^{-1}$), and L$_{UV}$ is the intrinsic (dust-corrected) non-ionising UV luminosity (in erg s$^{-1}$ Hz$^{-1}$), typically measured at 1500\AA\, as we have done. 

This follows the assumption that most of the ionising emissions from massive stars is absorbed by their birth cloud nebulae, emitting a significant amount of H$\alpha$ among other common lines. This also assumes that no LyC escapes the nebula. 

\subsection{Determining $\mathrm{f_{esc}}$}\label{subsec:fesc}
Without an intrinsic understanding of the LyC photons produced and a direct measure of the escaping portion, determining $\mathrm{f_{esc}}$ comes down to empirical probes determined at low redshift \citep{Izotov2018,Izotov2022}. Applying these to higher redshift galaxies is scientifically ambiguous, as the environments from which the LyC escapes are distinct in more evolved systems (see ionisation vs density bounded nebula discussion in \citet{Zackrisson2013} for an explanation of the screen and hole models). Using the relation from \cite{Chisholm2022}:
\begin{equation}
    f_{esc} = (1.3 \pm 0.6)\times{10}^{-4}\times10^{(-1.22\pm 0.1)\beta},
\end{equation}
we determine the UV slope ($\beta$) of the attenuated SED between 1500-2500\AA\, and derive the escape fraction ($\mathrm{f_{esc}^\beta}$) using this relation. 

We also introduce an approach based on a logical comprehension of the escape fraction similar in principle to previous attempts \citep{Oey1997,Papovich2026,Giovinazzo2026}. As described in section \ref{subsec:xion}, the calculation of $\mathrm{\xi_{ion}}$ relies on an assumption of $\mathrm{f_{esc}}$ = 0, as the dust corrected H$\alpha$ emission represents the absorbed LyC. Using the SED model, however, we can predict the total intrinsic LyC emission produced, and create a ratio of the H$\alpha$ emission to this. This ratio therefore represents the fraction of LyC absorbed by the nebula, so subtracting this result from 1 reflects the escaping fraction. As the H$\alpha$ probe is being used to indicate only the LyC absorption, this must be dust corrected as described in section \ref{subsec:xion} to derive $L_{H\alpha}^{CF00}$. 
\begin{equation}
    f_{esc}^{neb} = 1-\frac{N_{ion}^{abs}}{N_{ion}^{total}} = 1-\frac{7.28\times 10^{11} \times L_{H\alpha}^{CF00}}{N^{Model}_{ion}} 
\end{equation}
Where 7.28$\times 10^{11}$ is the linear scaling factor from \cite{2006agna.book.....O}, $L_{H\alpha}^{CF00}$ refers to the dust corrected H$\alpha$ luminosity (erg s$^{-1}$) and N$_{ion}^{Model}$ is the total calculated ionising emission from the \texttt{BEAGLE} model. We then consider dust attenuation of both the H$\alpha$ emission and direct absorption of LyC by dust in the galaxy separately. We use the CF00 dust attenuation model on the region below 912\AA\, integrating the relation so that we may account for any direct LyC absorption by dust. We assume a flat LyC spectrum (constant L$_\lambda$) so that: 
\begin{equation}
    f_{esc}^{dust} = \frac{\int^{912}_{100} L_\lambda e^{-\tau(\lambda)}d\lambda }{\int^{912}_{100} L_\lambda d\lambda} = \frac{1}{912-100}\int^{912}_{100}e^{-\tau(\lambda)}d\lambda
\end{equation}
and discuss the implications of this choice in Appendix \ref{AppendixA}. CF00 treats interstellar medium (ISM) and birth cloud (BC) attenuation separately as $\tau_\lambda = \tau_\lambda^{ISM} + \tau_\lambda^{BC}$ where $\tau_\lambda^x = \tau_V^x (\frac{\lambda}{5500})^{-n}$. The $\tau_\lambda^{BC}$ affects only massive stars still within their nebula which experience a stronger ($n=1.3$) attenuation, while $\tau_\lambda^{ISM}$ affects all stars with a weaker ($n=0.7$) attenuation. We combine these to derive:
\begin{equation}
    \tau(\lambda) = \left(\frac{A_V}{1.086}\right) \times \left(\frac{600}{5500}\right)^{-0.7}\times \left(\frac{\lambda}{5500}\right)^{-1.3},
\end{equation}  
which uses a reference wavelength of 600\AA\, for the ISM component rather than integrating over the LyC wavelength range, and uses A$_V$ = 1.086$\tau$ as in \cite{Siebenmorgen2014}.  This gives us the final equation for the LyC dust-escape fraction as:

\begin{align}
f_{\text{esc}}^{\text{dust}} = &\frac{1}{912 - 100} \times \nonumber\\\int_{100}^{912} &\exp\left[ -\left( \frac{A_V \times (600/5500)^{-0.7}}{1.086} \right) \left( \frac{\lambda}{5500} \right)^{-1.3} \right] \, d\lambda,
\end{align}
where the final $f_{esc}$ is just:

\begin{equation}
    f_{esc} = f_{esc}^{nebula} \times f_{esc}^{dust}
\end{equation}
We further derive an $f_{esc}^{Model}$ and an $f_{esc}^{H\alpha}$ which just uses either the \texttt{BEAGLE} model H$\alpha$ or the OutThere spectral H$\alpha$ for the f$_{esc}^{neb}$ calculation. 

We note that we explicitly use an SED model with an intrinsic $f_{\mathrm{esc}} = 0$ when deriving nebular emission properties, such that all ionising photons are assumed to be processed locally within the nebula. This minimises additional uncertainties associated with poorly constrained LyC escape and avoids artificially suppressing or modifying the predicted nebular line strengths through model-dependent escape assumptions. Allowing non-zero $f_{\mathrm{esc}}$ may introduce additional degeneracy between ionising photon production and nebular line strength, as escaping LyC photons no longer contribute to recombination emission. By fixing $f_{\mathrm{esc}} = 0$, the inferred nebular properties are driven primarily by the observed photometry rather than assumptions about LyC leakage.

\subsection{Quality Control}
\texttt{BEAGLE} provides a number of diagnostics to check the convergence of the walkers on a solution, which functions best on a case-by-case basis \citep{Chevallard2016}. However, when assessing the quality of the fit for a large sample this becomes impractical. We therefore employed the measured H$\alpha$ emission line from the OutThere spectra and measured the fractional difference to the modelled value as:
\begin{equation}
    \mathrm{\frac{H\alpha^{Model}-H\alpha}{H\alpha}}
\end{equation}
as a separate diagnostic of the emission line fit quality. The distribution of this diagnostic can be found in Figure \ref{fig:QC} and as the colour bar in each of the results. We found that some galaxies derived H$\alpha$ values that were more than 2$\times$ as large as the spectral H$\alpha$, which are represented as cyan triangles in our results (15/230). The 16-84 percentile range for this sample diagnostic is -0.72 to 0.71 with a median of -0.18, indicating only a minor underestimation bias for the majority of the sample.
The reduced $\chi^2$ in Figure \ref{fig:QC} is defined first as:
\begin{align}
    \mathrm{\chi^2_{raw}=\sum_{filters}(\frac{flux\_obs - flux\_model}{obs\_err})^2}\\ \mathrm{|{\frac{flux\_obs - flux\_model}{obs\_err}|}<3}
\end{align} where any residuals $>3$ are set to 0 to avoid dismissing the SED on one poorly fit filter or filters containing emission line contamination. This is then reduced using 
\begin{equation}
    \mathrm{\chi^2 = \frac{\chi^2_{raw}}{(N_{filters}-1)}}
\end{equation} where $\mathrm{N_{filters}}$ equalled the number of non-zero photometric filters for that source.

For each plot throughout this paper, we colour the data-points by their agreement with the nebular line estimation and make them less opaque if the reduced $\chi^2$ is large ($\sim1.5)$. Binning for the sake of creating the overplotted sample medians is performed such that each bin contains a similar number of points along the x-axis parameter for each plot. This was chosen rather than an equidistant binning approach as this would over-represent extreme values. When binning parameters we exclude values represented as cyan triangles.     

\subsection{Sample Introduction}
Refer to Table \ref{tab:physical_params} for the median physical properties and parameter ranges of the sample. Using the main-sequence relation defined by \cite{Speagle14} (their Equation 28), we find that only $\sim$15\% of galaxies lie within the expected main-sequence limits (Figure \ref{fig:MainSequence}). Defining recent burst activity as $\mathrm{SFR_{10}/SFR_{100} > 9}$, such that the recent star formation rate exceeds the estimated 100 Myr average by a factor of nine, yields $\sim$24\% of the sample. Conversely, defining declining or quiescent star formation as $\mathrm{SFR_{10}/SFR_{100} < 1}$ yields $\sim$20\% of galaxies.

The range of $\mathrm{M_{UV}}$ values indicates that the sample is generally brighter than those typically studied in the high-redshift literature. Additionally, a significant fraction of the sample ($\sim$23\%) would be classified as extreme [OIII]$\lambda5007$ equivalent width emitters, with $\mathrm{EW([OIII]\lambda5007) > 400,\AA}$ (Figure \ref{fig:O3Ha}; \citealt{Jaiswar2024}). The extreme [OIII]$\lambda5007$ and H$\alpha$ equivalent widths present in the sample allow us to probe the applicability of the $\mathrm{\xi_{ion}}$ relations derived by \cite{Tang2019} and \cite{Boyett2024} in regimes of extreme line emission.

\begin{table}
\centering
\caption{Physical parameter median, 1$\sigma$ and sample range for our final sample of 230 galaxies.}
\label{tab:physical_params}
\begin{tabular}{lcc}
\toprule
Parameter & Median & Range \\
\midrule
$\mathrm{log_{10}(M_\odot/M_*)}$ & $9.33^{+0.86}_{-0.74}$ & 7.22 , 11.41\\
$\mathrm{SFR_{100}/(M_*/yr)}$ & $1.36^{+6.08}_{-0.90}$ & 0.135 , 320\\
$\mathrm{M_{UV}}$ & $ -19.7^{-1.6}_{+0.9}$ & -25.9 , -17.3\\
$\mathrm{[OIII]5007}$\AA\,EW & $160^{+473}_{-111}$& 0.8 , 3850\\
H$\alpha$\AA\,EW & $287^{+533}_{-144}$& 12 , 11170\\
$\mathrm{A_V}$ & $0.14^{+0.37}_{-0.11}$ & 5.7$\times10^{-4}$ , 4.4\\
\bottomrule
\end{tabular}
\end{table}

\section{Results}

\subsection{$\mathrm{\xi_{ion}}$}
The ionising photon production efficiency is important to constrain for the purpose of understanding the ionising photon budget in the EoR. Deriving correlations to other physical parameters that are easier to constrain for large samples is vital to draw broader conclusions on the evolutionary history of the Universe in this epoch. Here we explore the relationship of $\mathrm{\xi_{ion}^{HII}}$ to physical parameters determined from \texttt{BEAGLE}.

\begin{table*}
\centering
\caption{$\xi_{\mathrm{ion}}$ (erg$^{-1}$ Hz) and f$\mathrm{_{esc}}$ (\%) measured in five bins.
For each parameter, bin centres are shown in the first row, followed by either $\xi_{\mathrm{ion}}$ or $\mathrm{f_{esc}}$. Bins are equal-population bin medians with the 16-84th percentile ranges reported.}
\label{tab:xi_ion_bins}

\begin{tabular}{l c c c c c}
\hline
Parameter 
& Bin 1 
& Bin 2 
& Bin 3 
& Bin 4 
& Bin 5 \\
\hline
UV slope  
& -1.85 & -1.62 & -1.49 & -1.37 & -0.68 \\
$\xi_{\mathrm{ion}}$
& 24.96$^{+0.37}_{-0.37}$ 
& 24.94$^{+0.35}_{-0.19}$ 
& 24.88$^{+0.27}_{-0.22}$ 
& 24.94$^{+0.27}_{-0.28}$ 
& 25.05$^{+0.43}_{-0.46}$ \\

M$\mathrm{_{UV}}$  
& -21.87 & -20.46 & -19.68 & -19.12 & -18.58 \\
$\xi_{\mathrm{ion}}$
& 24.89$^{+0.47}_{-0.27}$ 
& 24.90$^{+0.18}_{-0.23}$ 
& 24.96$^{+0.38}_{-0.23}$ 
& 24.94$^{+0.36}_{-0.18}$ 
& 24.96$^{+0.41}_{-0.34}$ \\

log$_{10}$(M$_\odot$/M$_*$)   
& 8.31 & 8.97 & 9.33 & 9.74 & 10.46 \\
$\xi_{\mathrm{ion}}$
& 24.95$^{+0.29}_{-0.39}$ 
& 24.95$^{+0.41}_{-0.21}$ 
& 24.93$^{+0.31}_{-0.25}$ 
& 24.91$^{+0.09}_{-0.24}$ 
& 25.00$^{+0.46}_{-0.37}$ \\

log$_{10}$(sSFR/yr) 
& -9.77 & -9.28 & -8.90 & -8.51 & -7.53 \\
$\xi_{\mathrm{ion}}$
 & 24.91$^{+0.35}_{-0.24}$ 
 & 24.96$^{+0.33}_{-0.20}$ 
 & 24.90$^{+0.40}_{-0.31}$ 
 & 24.88$^{+0.25}_{-0.23}$ 
 & 25.00$^{+0.38}_{-0.36}$ \\

[OIII]5007 EW 
& 31.46 & 83.10 & 160.45 & 288.40 & 897.74 \\
$\xi_{\mathrm{ion}}$
& 24.86$^{+0.49}_{-0.25}$ 
& 24.87$^{+0.17}_{-0.29}$ 
& 24.86$^{+0.47}_{-0.19}$ 
& 25.00$^{+0.22}_{-0.24}$ 
& 25.08$^{+0.34}_{-0.20}$ \\

H$\alpha$ EW  
& 109.63 & 196.82 & 286.64 & 460.57 & 1077.85 \\
$\xi_{\mathrm{ion}}$
& 24.67$^{+0.35}_{-0.25}$ 
& 24.86$^{+0.13}_{-0.21}$ 
& 24.88$^{+0.18}_{-0.12}$ 
& 25.01$^{+0.26}_{-0.14}$ 
& 25.35$^{+0.23}_{-0.25}$ \\

z 
& 1.705 & 1.783 & 1.939 & 2.123 & 2.292 \\
f$\mathrm{_{esc}^{\beta}}$/\%
& 0.710$^{+1.053}_{-0.358}$ \%
& 0.868$^{+1.128}_{-0.533}$ \%
& 1.155$^{+1.217}_{-0.795}$ \%
& 0.794$^{+0.628}_{-0.480}$ \%
& 0.871$^{+1.375}_{-0.680}$ \%\\

z 
& 1.707 & 1.779 & 1.955 & 2.140 & 2.294 \\
f$\mathrm{_{esc}^{H\alpha}}$/\%
& 0.067$^{+15.777}_{-0.067}$ \%
& 0.089$^{+17.978}_{-0.089}$ \%
& 0.200$^{+9.041}_{-0.200}$ \%
& 0.006$^{+1.852}_{-0.006}$ \%
& 0.002$^{+2.069}_{-0.002}$ \%\\

z 
& 1.705 & 1.783 & 1.939 & 2.123 & 2.292 \\
f$\mathrm{_{esc}^{Model}}$/\%
& 0.682$^{+23.964}_{-0.682}$ \%
& 0.083$^{+20.569}_{-0.083}$ \%
& 0.050$^{+18.578}_{-0.050}$ \%
& 0.005$^{+2.210}_{-0.005}$ \%
& 0.002$^{+1.050}_{-0.002}$ \%\\
\hline
\end{tabular}
\end{table*}

\subsubsection{UV parameters}
The UV slope is calculated as the slope of UV luminosity between 1500-2500\AA\, in the  modelled SED from \texttt{BEAGLE}. (Figure \ref{fig:xion-bmuv}). We report the median $\mathrm{\xi_{ion}^{HII}}$ of each UV slope bin from bluest to reddest in Table \ref{tab:xi_ion_bins}. The assumed dependence of $\mathrm{\xi_{ion}}$ on the massive stellar population and the expected correlation of a young stellar population to a bluer UV slope suggest that this relationship should too, be negatively correlated. Bluer galaxies should be more efficient ionising photon producers. However, for our sample, we do not find a significant Spearman correlation (p$>$0.05). We note, however, that when excluding data-points within the reddest bin, we do find a significant positive correlation ($\rho$ =0.28, p$<7.1\times10^{-4}$). \cite{Harshan2024} and \cite{Lam2019} similarly find no correlation, also using a smaller dynamical range of UV slopes, while \cite{Castellano2023} and \cite{Bouwens2015} find a negative correlation over a wider range. The difficulty with associating the UV slope with stellar age comes with its degeneracy with dust attenuation. In our sample, the reddest galaxies also have the largest A$_V$ (see Figure \ref{fig:QC}).  

We calculate the median of each M$\mathrm{_{UV}}$ bin in Table \ref{tab:xi_ion_bins}. We report no significant Spearman correlation (p$>$0.05) though this hinges on the inclusion of the brightest bin ($\rho$ =0.19, p$<$0.02 without it). \cite{Harshan2024} finds a steep positive correlation at z$>5$ and suggests that fainter galaxies are more efficient producers of ionising emission, though for our sample at z$\sim$1.5 this does not appear to be the case. In both the case of the $\beta$ and $\mathrm{M_{UV}}$ correlations, removing data-points within the reddest/brightest bin improves the significance of the correlation significantly.

\subsubsection{Stellar Parameters}
The stellar mass (log$_{10}$(M$_*$/M$_\odot$)) and specific star formation rate (log$_{10}$(sSFR/yr$^{-1}$)) are each derived from the \texttt{BEAGLE} best fit SED (Figure \ref{fig:xion-bmuv}). We derive bin-median stellar masses from the least to most massive in Table \ref{tab:xi_ion_bins}. We do not find a significant Spearman correlation with this parameter (p$>$0.05), suggesting that more massive galaxies do not produce less ionising emission per non-ionising UV emission. This is unexpected as the $\mathrm{\xi_{ion}}$ dependence on a young stellar population would be linked to its mass via the age-mass relation. We note, however, that this correlation is improved when removing data-points within the most massive bin with sample variance roughly 1.6$\times$ that of the other bins. \cite{Harshan2024, Lam2019, Castellano2023} find negative slopes with a similarly large intrinsic scatter, albeit with a stronger overall ionising efficiency for a less massive, higher redshift sample.

We compute the sSFR bin-medians in Table \ref{tab:xi_ion_bins} from most negative to most positive. We do not find a significant Spearman correlation in these parameters. Similar works by \cite{Harshan2024} also using the SED derived SFR, finds only a weak positive correlation at z$>$5 despite a strong theoretical support to this relationship. Proportionally high star formation rates indicate young stellar populations that should have a large contribution to the ionising photon production. \cite{Castellano2023} finds this correlation to be strong in their 2$<z<$5 sample, however, both of these works claim increasing scatter and weaker correlation toward the lower sSFR end where the majority of our sample lies (sSFR $<$ -8). This correlation may therefore hold better for starburst galaxies at higher redshifts than our sample.    
\begin{figure}[h!]
    \centering
    \includegraphics[width=0.8\linewidth]{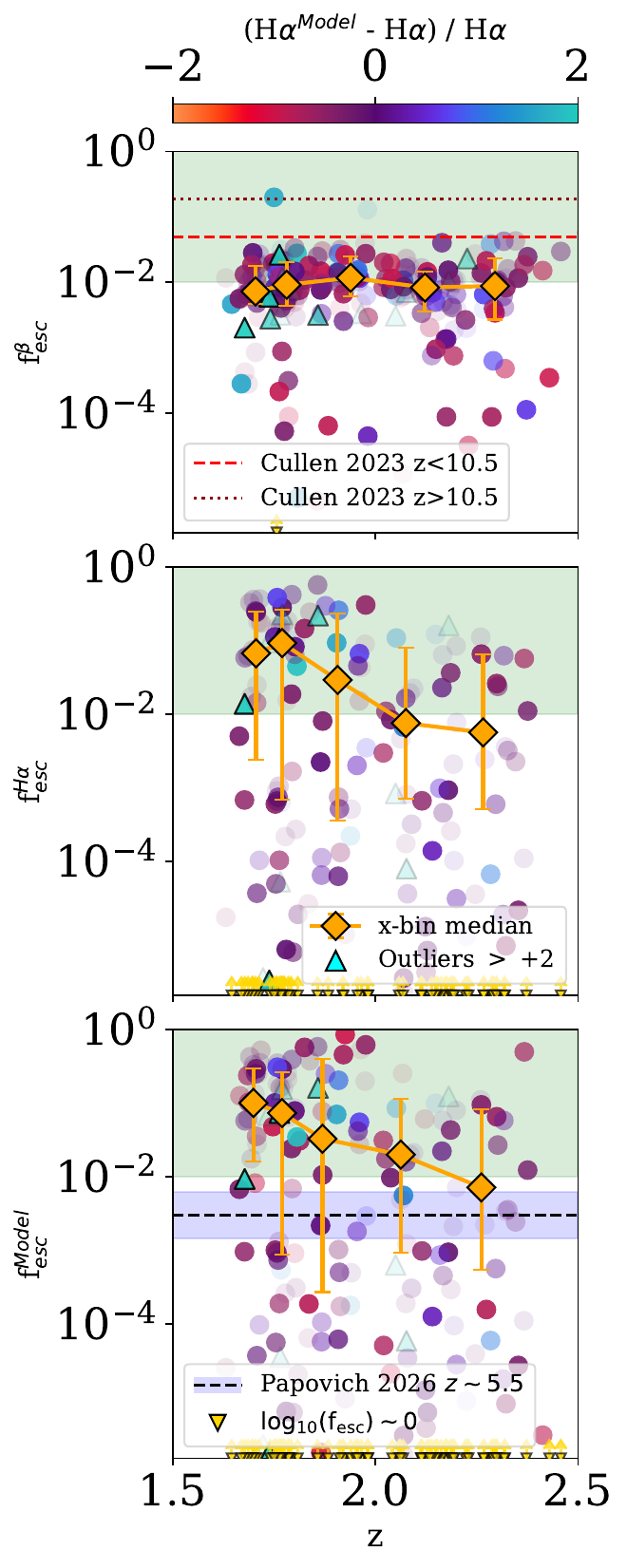}
    \caption{Escape fraction of Lyman Continuum radiation calculated using (Top) the UV slope, and the theoretical approach described in section \ref{subsec:fesc}  using spectral H$\alpha$ (Middle) and using modelled H$\alpha$ (Bottom). Region above 1\% escape fraction indicated by the green shaded region. Opacity of data-points inversely proportional to \texttt{BEAGLE} SED reduced $\chi^2$ so that better fits are more opaque. Orange diamonds correspond to equal-population bin medians, and their error-bars reflect the 16-84th percentile ranges. \cite{Cullen2024} sample averages for z$<$10.5 and  z$>$10.5 shown in the UV slope panel using the same method. \cite{Papovich2026} uses the SED modelled $\mathrm{f_{esc}}$ which we include in the bottom panel for comparison. Our sample at z$\sim$2 has a much lower $\mathrm{f_{esc}^\beta}$ and a much more scattered $\mathrm{f_{esc}^{Model}}$ than the comparison samples.}
    \label{fig:fescredshift}
\end{figure}

\begin{figure*}
    \centering
    \includegraphics[width=0.95\linewidth]{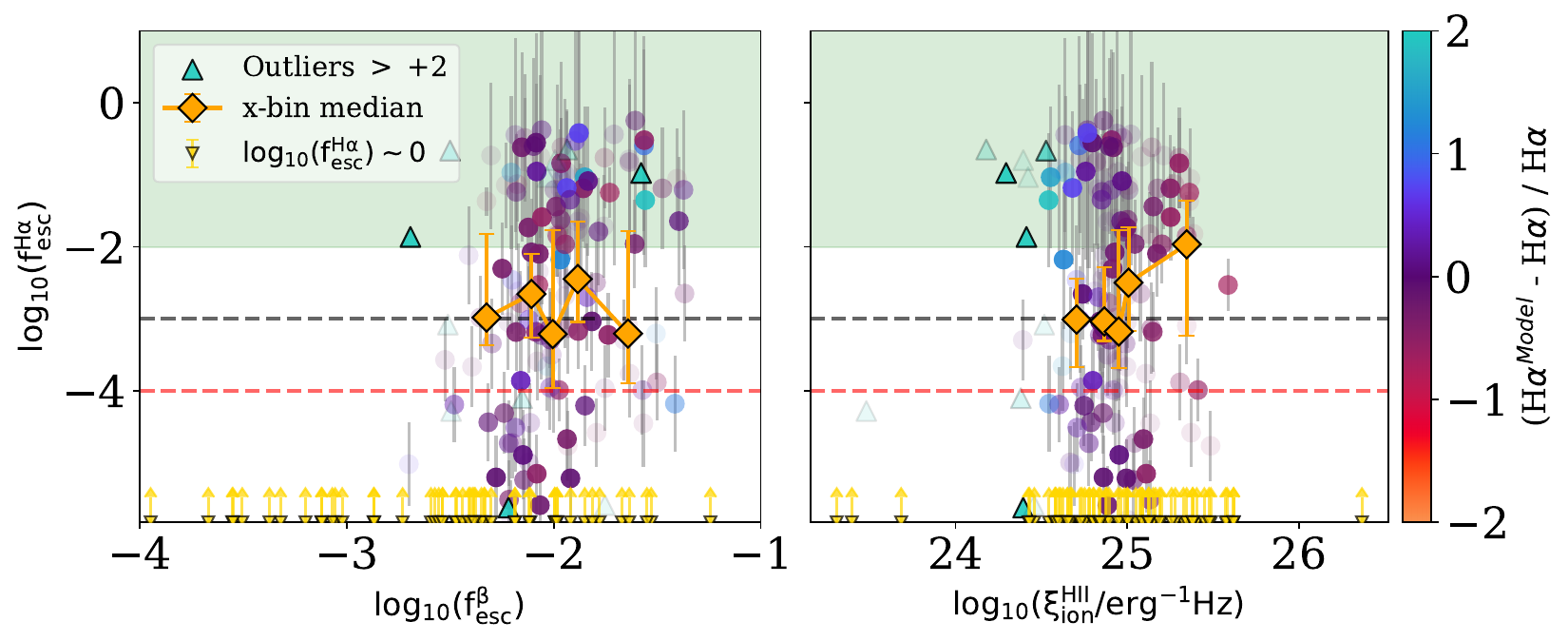}
    \caption{(Left) $\mathrm{f_{esc}}$ determined from the \cite{Chisholm2022} relation ($\mathrm{f_{esc}^\beta}$) vs determined from our spectroscopic derivation ($\mathrm{f_{esc}^{H\alpha}}$). (Right) $\mathrm{\xi_{ion}^{HII}}$ vs the $\mathrm{f_{esc}^{H\alpha}}$. Lower limits indicate $\mathrm{f_{esc}^{H\alpha}}$ values below the 30th percentile ($<1.5\times10^{-6}$). Orange diamonds correspond to equal-population bin medians (excluding limits), and their error-bars reflect the 16-84th percentile ranges. Green shaded region indicates $\mathrm{f_{esc}^{H\alpha}>1\%}$ (64/199 sources) while black and red dotted lines represent $\mathrm{f_{esc}^{H\alpha} = 0.1\%}$ and $\mathrm{f_{esc}^{H\alpha} = 0.01\%}$ respectively. We do not find a significant correlation with either parameter.}
    \label{fig:fescionising}
\end{figure*}

\begin{figure}
    \centering
    \includegraphics[width=0.99\linewidth]{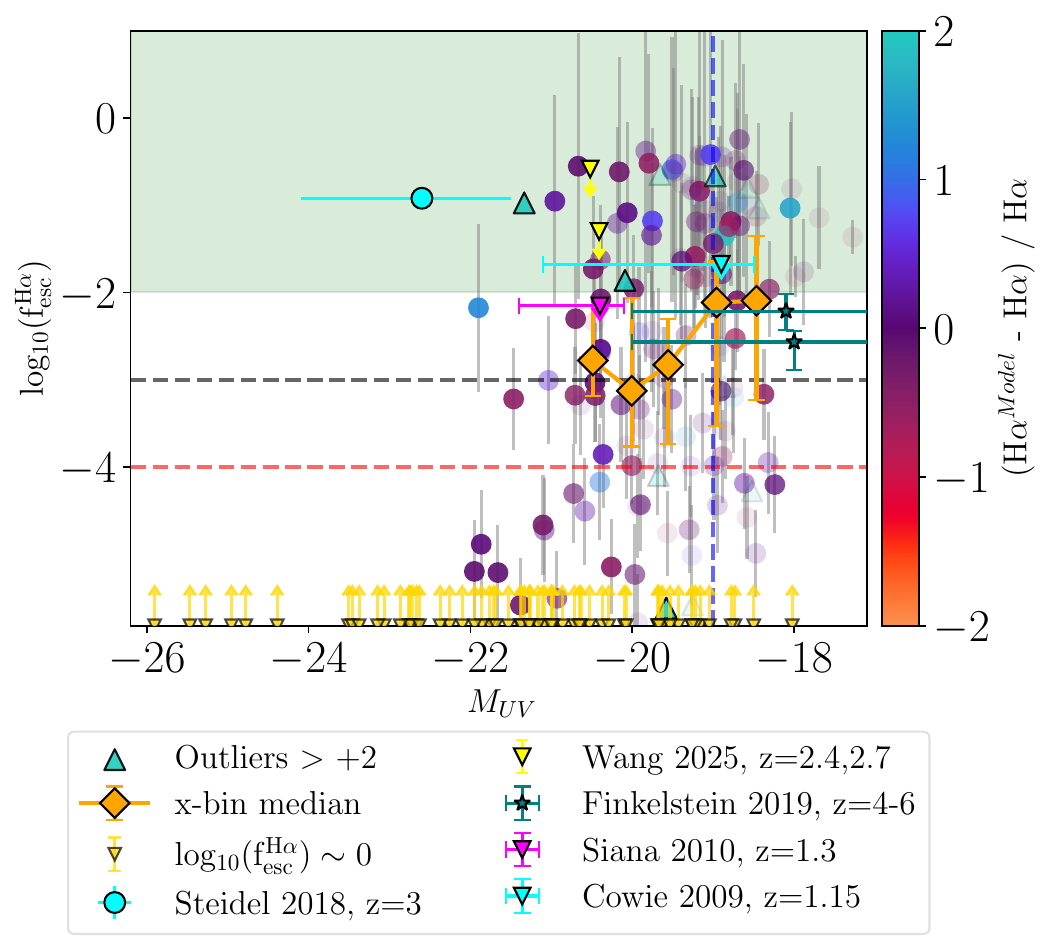}
    \caption{$\mathrm{M_{UV}}$ vs $\mathrm{f_{esc}}$ determined from our spectroscopic derivation ($\mathrm{f_{esc}^{H\alpha}}$). Lower limits indicate $\mathrm{f_{esc}^{H\alpha}}$ values below the 30th percentile ($<1.5\times10^{-6}$). Orange diamonds correspond to equal-population bin medians (excluding limits), and their error-bars reflect the 16-84th percentile ranges. Green shaded region indicates $\mathrm{f_{esc}^{H\alpha}>1\%}$ (64/199 sources) while black and red dotted lines represent $\mathrm{f_{esc}^{H\alpha} = 0.1\%}$ and $\mathrm{f_{esc}^{H\alpha} = 0.01\%}$ respectively. Blue vertical line indicates the $\mathrm{M_{UV} = -19}$ cutoff. We compare our results to LyC selected studies \citep{Steidel2018, Wang_2025} as well as non-selected stacks \citep{Cowie2009,Siana_2010} and the \cite{Finkelstein2019} simulation at z=4-6. We find significant positive correlations between the $\mathrm{f_{esc}^{H\alpha}}$ and $\mathrm{M_{UV}}$ parameters ($\rho >$ 0.22) suggesting that fainter $\mathrm{M_{UV}}$ galaxies have larger escape fractions.}
    \label{fig:fescMuv}
\end{figure}

\begin{figure*}
    \centering
    \begin{subfigure}{\linewidth}
        \centering
        \includegraphics[width=0.95\linewidth]{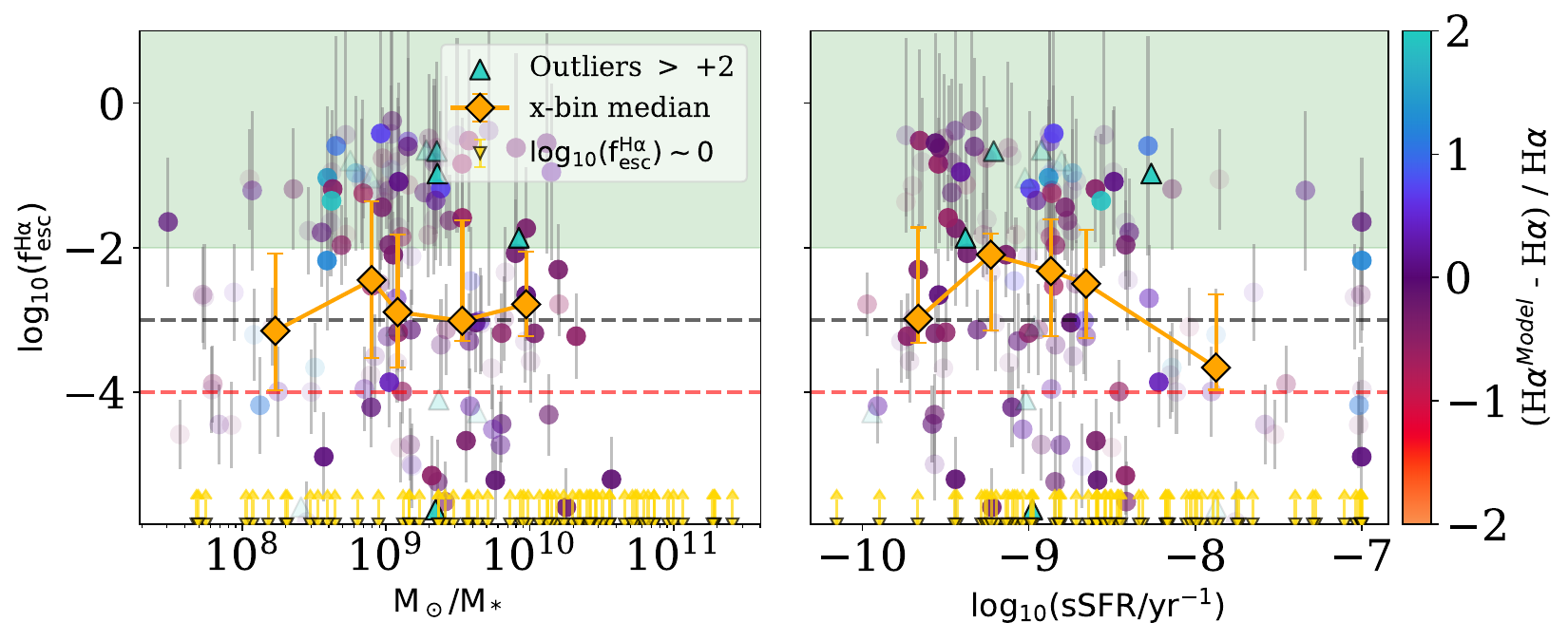}
        \caption{}
        \label{fig:fesc3}
    \end{subfigure}

    \vspace{0.8em}

    \begin{subfigure}{\linewidth}
        \centering
        \includegraphics[width=0.95\linewidth]{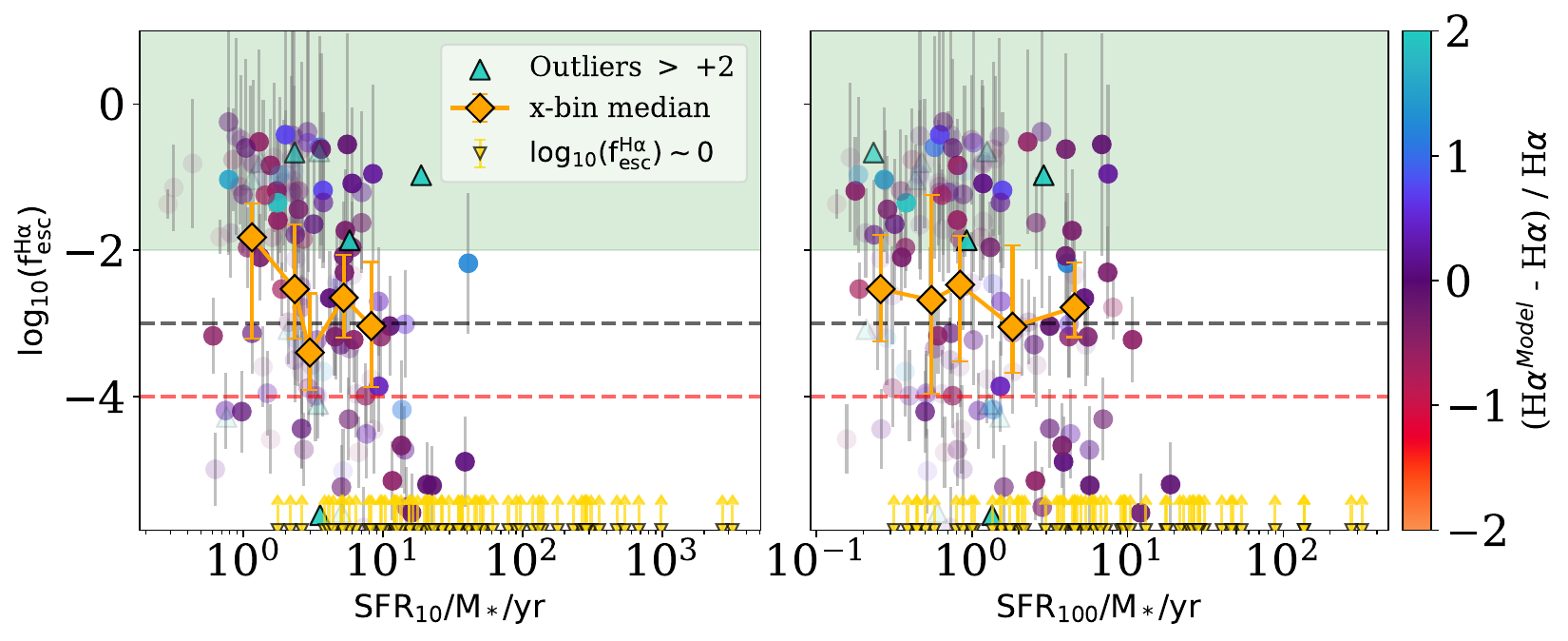}
        \caption{}
        \label{fig:fesc4}
    \end{subfigure}
    \caption{a) Left) Stellar mass vs $\mathrm{f_{esc}^{H\alpha}}$ and Right) Mass normalised star formation rate (sSFR) vs $\mathrm{f_{esc}^{H\alpha}}$. b) Left) Short timescale averaged SFR (10 Myr) and Right) 100 Myr timescale averaged SFR vs $\mathrm{f_{esc}^{H\alpha}}$. Lower limits indicate $\mathrm{f_{esc}^{H\alpha}}$ values below the 30th percentile ($<1.5\times10^{-6}$). Orange diamonds correspond to equal-population bin medians (excluding limits), and their error-bars reflect the 16-84th percentile ranges. Green shaded region indicates $\mathrm{f_{esc}^{H\alpha}>1\%}$ (64/199 sources) while black and red dotted lines represent $\mathrm{f_{esc}^{H\alpha} = 0.1\%}$ and $\mathrm{f_{esc}^{H\alpha} = 0.01\%}$ respectively. We observe a negative correlation in the sSFR suggesting that galaxies with relatively high rates of star formation do not allow LyC escape. We observe negative correlations to only the short timescale SFR, indicating that more bursty star forming galaxies do not have large escape fractions.}
    \label{fig:fesc_physical}
\end{figure*}

\begin{figure*}
    \centering
    \includegraphics[width=0.95\linewidth]{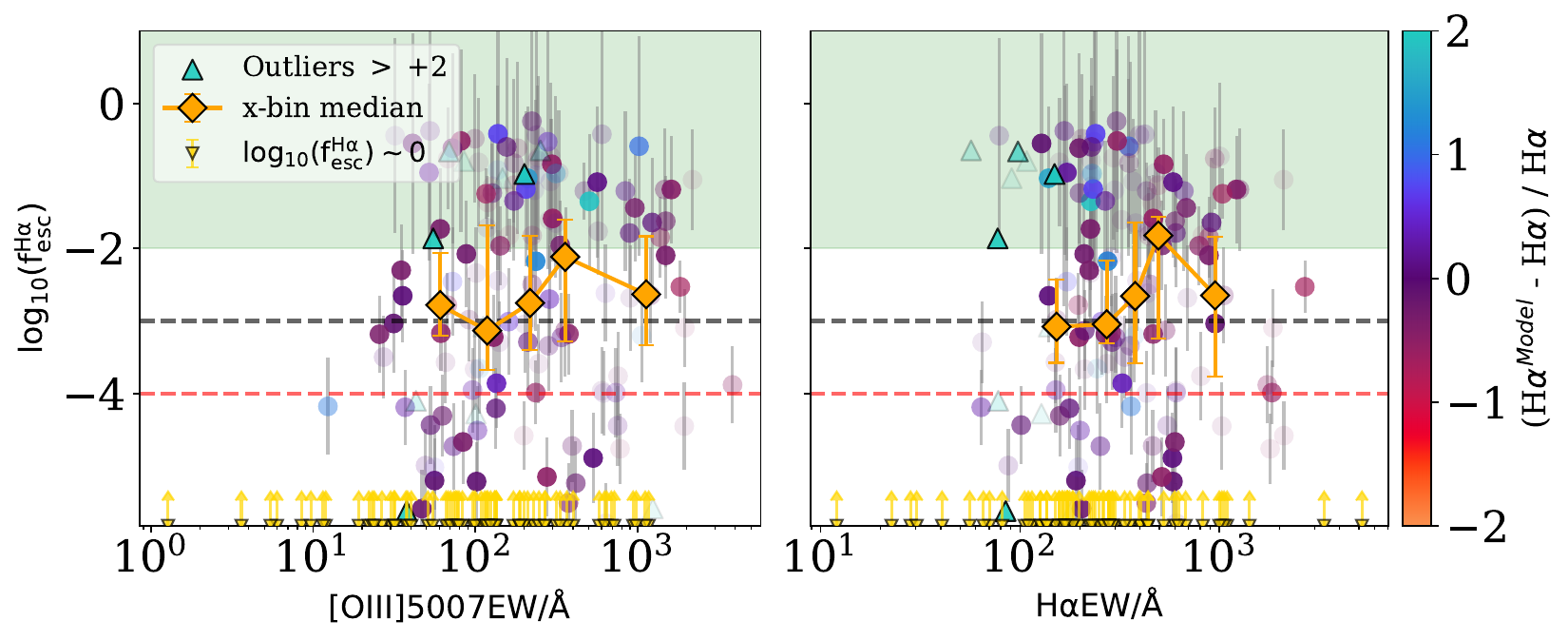}
    \caption{Left) $\mathrm{[OIII]}\,5007\,\lambda\text{\AA}$EW and Right) $\mathrm{H\alpha}\,\text{\AA}\,$EW vs $\mathrm{f_{esc}^{H\alpha}}$. Lower limits indicate $\mathrm{f_{esc}^{H\alpha}}$ values below the 30th percentile ($<1.5\times10^{-6}$). Orange diamonds correspond to equal-population bin medians (excluding limits), and their error-bars reflect the 16-84th percentile ranges. Green shaded region indicates $\mathrm{f_{esc}^{H\alpha}>1\%}$ (64/199 sources) while black and red dotted lines represent $\mathrm{f_{esc}^{H\alpha} = 0.1\%}$ and $\mathrm{f_{esc}^{H\alpha} = 0.01\%}$ respectively. We do not observe significant correlations to the $\mathrm{[OIII]}\,5007\,\lambda\text{\AA}$, while the $\mathrm{H\alpha}\,\text{\AA}\,$EW appears to be negatively correlated with the ionising escape. This may be an underlying degeneracy with bursty star formation.}
    \label{fig:fesco3ha}
\end{figure*}

\subsubsection{Correlations to Equivalent Widths}
The [OIII]5007 and H$\alpha$ EW are derived from the OutThere NIRISS spectra (Figure \ref{fig:o3}). We report the bin medians from lowest to highest [OIII]5007\AA\, EW in Table \ref{tab:xi_ion_bins} and find a significant Spearman correlation (slope = 0.16, intercept = 24.62, Spearman rank = 0.26 p$<$5.0$\times 10^{-5}$). The [OIII]5007 emission line is associated with star forming regions, and theoretically this suggests large [OIII]5007 EW should correlate with young stellar populations which our results corroborate. We highlight the extreme [OIII]5007 region and find these sources exhibiting the largest $\mathrm{\xi_{ion}^{HII}}$.  \cite{Tang2019} and \cite{Boyett2024} find a strict correlation between $\mathrm{\xi_{ion}^{HII}}$ measurements and the dust-corrected, \texttt{BEAGLE} SED L$_{UV}$ which contains both stellar and nebular contributions. These appear to be dependent on the sample redshift \citep{Tang2019,Boyett2024}. We find that the correlations hold within 3$\sigma$ for our sample spanning a wider redshift range and constraining the lower dynamical range of [OIII]5007 EW$<225$. \cite{Tang2019} predicted deviations below [OIII]5007 EW $<$225, and we do find a larger scatter below this limit (see Table \ref{tab:xi_ion_bins}), which reduces the measured slope and Spearman rank of our sample by comparison to a relation drawn only considering values above this (slope = 0.413, intercept = 23.94, Spearman rank = 0.352, p$<$6.3$\times10^{-4}$). 

The H$\alpha$ EW is itself correlated with the H$\alpha$ flux used to derive $\mathrm{\xi_{ion}^{HII}}$ as a tracer of LyC emission. We calculate the bin averages of this relation in Table \ref{tab:xi_ion_bins} from lowest to highest EW. We also report a clear positive monotonic relationship (slope = 0.53, intercept = 23.65, Spearman rank = 0.65, p$<$6.4$\times10^{-29}$) that matches that found in \cite{Tang2019} within 1$\sigma$. \cite{Harshan2024} finds a similar slope (0.7) and Spearman coefficient (0.6) at z$>$5.

\subsection{$\mathrm{f_{esc}}$}
The escape fraction of ionising radiation is necessary to constrain alongside the ionising photon production efficiency to determine the ionising budget per galaxy. The challenge with estimating this value beyond the Local Universe is the direct detection of LyC which is readily absorbed by the IGM even at intermediate redshifts \citep{Inoue2014a}. Here we explore the correlations of our escape fraction estimates to physical parameters determined from \texttt{BEAGLE} to test our methodology and recover potential correlations. Of the 233 galaxies in our sample, 199 have measurable f$\mathrm{_{esc}}$ values after excluding systems with physically inconsistent spectral and SED-derived properties. Specifically, we remove galaxies where the total ionising photon budget, N$\mathrm{_{ion}}$, inferred from the SED model is lower than that implied by the dust-corrected H$\alpha$ luminosity, as well as cases yielding non-physical escape fractions (f$\mathrm{_{esc}}<0$ or f$\mathrm{_{esc}}>1$). 

The majority of the following analysis is performed using the method described in section \ref{subsec:fesc} which derives the escape fraction we refer to as f$_{esc}^{H\alpha}$. We denote the UV slope derived $\mathrm{f_{esc}}$ as f$_{esc}^\beta$ and distinguish $\mathrm{f_{esc}}$ calculated from the spectral H$\alpha$ from the SED model H$\alpha$ as f$_{esc}^{H\alpha}$ and f$_{esc}^{Model}$ respectively. 

Our methodology is not directly comparable to much of the existing literature in this field. We do not treat the $\mathrm{f_{esc}}$ as a free parameter in the SED model fit. The closest comparisons are studies that derive $\mathrm{f_{esc}}$ directly from SED models, where the escape fraction is incorporated into the fitting procedure \citep{Giovinazzo2026,Papovich2026}. We instead use the SED model to inform the total theoretical ionising budget and the spectroscopic H$\alpha$ (dust corrected using the SED model) as our probe of its absorption by the nebula. Combined with the dust absorption estimation (using the SED model), this results in a large dynamical range of possible f$_{esc}^{H\alpha}$ compared to methods relying on selected priors. We therefore treat extremely low values of f$_{esc}^{H\alpha}$ as lower limits, and report correlations both including and excluding these systems.  

Studies deriving SED-modelled $\mathrm{f_{esc}}$ typically impose lower priors of 0.1 - 0.01\%. However, the physical plausibility of smaller escape fractions is not constrained by the practical limits adopted in SED fitting. We therefore adopt 0.01\% as our fiducial lower limit, and vary this between 0.1\% and no limit to verify that the qualitative behaviour of the correlations remains unchanged when varying the floor.

\begin{table*}
\centering
\caption{
Spearman correlation coefficients ($\rho$) and associated p-values between ${\mathrm{f_{esc}^{H\alpha}}}$ and galaxy properties for different lower escape fraction thresholds. cells reflecting correlations with p-value$>$0.05 are highlighted in green, while cells highlighted in red fall above this. 
}
\label{tab:fesc_threshold_corr}

\begin{tabular}{l cc cc cc}
\toprule
${\mathrm{f_{esc}^{H\alpha}}}$ Lower Limit
& \multicolumn{2}{c}{$0.1\%$}
& \multicolumn{2}{c}{$0.01\%$}
& \multicolumn{2}{c}{No Threshold} \\

\cmidrule(lr){2-3}
\cmidrule(lr){4-5}
\cmidrule(lr){6-7}

Parameter
& $\rho$ & p-value
& $\rho$ & p-value
& $\rho$ & p-value \\

\midrule

Redshift ($z$)
& \cellcolor{green!15}-0.38
& \cellcolor{green!15}$<5\times10^{-4}$
& \cellcolor{green!15}-0.28
& \cellcolor{green!15}$<3\times10^{-3}$
& \cellcolor{green!15}-0.21
& \cellcolor{green!15}$<3\times10^{-3}$
\\

$\mathrm{f_{esc}^{\beta}}$
& \cellcolor{red!15}
& \cellcolor{red!15}
& \cellcolor{red!15}
& \cellcolor{red!15}
& \cellcolor{green!15}0.53
& \cellcolor{green!15}$<<1\times10^{-6}$
\\

${\mathrm{f_{esc}^{Model}}}$
& \cellcolor{green!15}0.90
& \cellcolor{green!15}$<<1\times10^{-6}$
& \cellcolor{green!15}0.96
& \cellcolor{green!15}$<<1\times10^{-6}$
& \cellcolor{green!15}0.99
& \cellcolor{green!15}$<<1\times10^{-6}$
\\

$\log_{10}(\xi_{\mathrm{ion}}^{HII})$
& \cellcolor{red!15}
& \cellcolor{red!15}
& \cellcolor{red!15}
& \cellcolor{red!15}
& \cellcolor{red!15}
& \cellcolor{red!15}
\\

$\log_{10}(\xi_{\mathrm{ion}}^{\mathrm{BEAGLE}})$
& \cellcolor{green!15}-0.31
& \cellcolor{green!15}$<5\times10^{-3}$
& \cellcolor{green!15}-0.32
& \cellcolor{green!15}$<<1\times10^{-6}$
& \cellcolor{green!15}-0.45
& \cellcolor{green!15}$<<1\times10^{-6}$
\\

$\log_{10}(M_{\star}/M_{\odot})$
& \cellcolor{red!15}
& \cellcolor{red!15}
& \cellcolor{red!15}
& \cellcolor{red!15}
& \cellcolor{green!15}-0.37
& \cellcolor{green!15}$<<1\times10^{-6}$
\\

$\log_{10}(\mathrm{SFR}_{10})$
& \cellcolor{green!15}-0.34
& \cellcolor{green!15}$<3\times10^{-3}$
& \cellcolor{green!15}-0.40
& \cellcolor{green!15}$<2\times10^{-5}$
& \cellcolor{green!15}-0.74
& \cellcolor{green!15}$<<1\times10^{-6}$
\\

$\log_{10}(\mathrm{sSFR})$
& \cellcolor{red!15}
& \cellcolor{red!15}
& \cellcolor{green!15}-0.23
& \cellcolor{green!15}$<2\times10^{-2}$
& \cellcolor{green!15}-0.31
& \cellcolor{green!15}$<7\times10^{-6}$
\\

$\mathrm{EW}([\mathrm{OIII}]\lambda5007)$
& \cellcolor{red!15}
& \cellcolor{red!15}
& \cellcolor{red!15}
& \cellcolor{red!15}
& \cellcolor{green!15}0.27
& \cellcolor{green!15}$<2\times10^{-4}$
\\

$\mathrm{EW}(\mathrm{H\alpha})$
& \cellcolor{green!15}-0.25
& \cellcolor{green!15}$<3\times10^{-2}$
& \cellcolor{red!15}
& \cellcolor{red!15}
& \cellcolor{green!15}0.17
& \cellcolor{green!15}$<2\times10^{-2}$
\\

$M_{\mathrm{UV}}$
& \cellcolor{green!15}0.22
& \cellcolor{green!15}$<5\times10^{-2}$
& \cellcolor{green!15}0.25
& \cellcolor{green!15}$<8\times10^{-3}$
& \cellcolor{green!15}0.58
& \cellcolor{green!15}$<<1\times10^{-6}$
\\

\bottomrule
\end{tabular}
\end{table*}

\subsubsection{redshift evolution}\label{subsec:redshift_evol}

Here we report the redshift evolution of the $\mathrm{f_{esc}}$ for this z$\sim$2 sample for each of the three derivation methods (Figure \ref{fig:fescredshift}). 

We report no $f^\beta_{esc}$ evolution across redshift (p$>0.05$). We find that our sample at z$\sim$2 is consistently below the escape fraction found in high redshift samples using the same technique \citep{Cullen2024} ($\mathrm{f_{esc}^\beta}$ = 5\% and $\mathrm{f_{esc}^\beta}$ = 19\% for the z$<$10.5 and z$>$10.5 ranges respectively). The $\mathrm{f_{esc}^\beta}$ derivation is proportional to the UV slope, and hence our redder sample lies below the much bluer, early-universe counterparts. Theoretically, the  $\mathrm{f_{esc}^\beta}$ is shown to increase with increasing redshift \citep{Chisholm2022} as it is inversely proportional to the UV slope which generally decreases with increasing redshift. As our sample spans a short redshift range it is likely that our dynamical range hinders the estimation of a significant correlation. 

We find that our theoretical escape fraction methods overlap with traditional SED fit $\mathrm{f_{esc}}$ \citep{Papovich2026} at high redshifts albeit with more scatter. This method is similar to our f$_{esc}^{Model}$ method in that all inputs are taken from the SED model, though our implementation instead uses a simplified escape-path framework rather than incorporating the escape fraction directly as a goodness-of-fit parameter within the SED fitting procedure. The large scatter observed in the theoretical escape fractions is likely a product of the many competing variables in this approach, relying on both modelled dust corrections, dust escape fractions and theoretical ionising photon production from the photo-ionisation model, as well as the spectroscopic H$\alpha$, each of which can vary substantially per galaxy. The bin median values, however, are within the theoretical ranges reported by \cite{Matthee2017,Grazian2017} and determined for galaxies in a similar $\mathrm{M_{UV}}$ range by \cite{Finkelstein2019}. \cite{Grazian2017} finds that at a slightly higher redshift of $3.27<z<3.4$ galaxies brighter than M$_{1500}\sim-19$ exhibit an upper limit of f$_{esc} < 1.7\%$. Studying H$\alpha$ and Ly$\alpha$ emitters at z$\sim$2.2, \cite{Matthee2017} finds f$_{esc} < 6.4\%$ by directly studying the LyC using GALEX/NUV and mean-stacking. We find that our theoretical methods have significant negative correlations to the redshift at each threshold (see Table \ref{tab:fesc_threshold_corr} for correlations to  f$_{esc}^{H\alpha}$) ($\rho$ = -0.35, p$<2\times10^{-3}$ at 0.1\%, $\rho$ = -0.27, p$<4\times10^{-3}$ at 0.01\% and $\rho$ = -0.21, p$<3\times10^{-3}$ with no lower limit for f$_{esc}^{Model}$).

This negative trend aligns with observations above $z>4.5$ made by \cite{Papovich2026} using SED modelling, suggesting that early universe galaxies contained much of their ionising photon production. As discussed, low redshift studies correlating the UV slope derived $\mathrm{f_{esc}^\beta}$ with redshift imply an increasing escape fraction with redshift \citep{Chisholm2022,Flury2022} and even with the simulation based escape fraction from \cite{Finkelstein2019} and with the results of the THESAN simulations \citep{thesan}. 

These contrary results should be considered in terms of the distinctions in their method of derivation. The THESAN simulations incorporate radiative transfer modelling that self-consistently calculates the ionisation state of hydrogen gas from the radiation field \citep{thesan-sim}. Studies such as \cite{Papovich2026} use the photo-ionisation modelling incorporated within the SED model to self-consistently determine the production and escape of ionising photons given the synthesised stellar population. Our approach first confirms the agreement of the spectroscopically observed nebular emission with the modelled spectrum and uses this to confirm the viability of the SED modelled stellar population before using the total modelled ionising budget to self-consistently determine the ionising escape using our two-layer escape model. By contrast, the \cite{Finkelstein2019} model is calibrated empirically using data from \cite{Finkelstein2016} at z$>$4, as is the \cite{Chisholm2022} model using empirical, low redshift LyC observations. 

\subsubsection{ionising parameters}\label{subsec:ionising_params}
We do not find significant correlations between f$_{esc}^{H\alpha}$ and $f^\beta_{esc}$ when excluding the lower limit values, though find a fairly strong correlation when including the full sample (see Table \ref{tab:fesc_threshold_corr} and Figure \ref{fig:fescionising}). Theoretically as these quantities are intended as descriptions of one phenomenon, their correlation is logical. As discussed in \ref{subsec:redshift_evol}, however, this is not the case as the derivation methods are incomparable. By contrast, the theoretically equivalent $\mathrm{f_{esc}^{Model}}$ approach correlates very strongly.

We find no significant correlations to the $\mathrm{\xi_{ion}^{HII}}$ despite both parameters relying on the same dust corrected H$\alpha$ flux. Considering that the ionising photon production efficiency is a reflection of the intrinsic ionising photon budget, and that our escape fraction model uses the SED model ionising photon budget, this leaves only the dust component of the escape fraction calculation as a distinguishing input. Intriguingly, the internal SED model $\mathrm{\xi_{ion}}$ does significantly correlate with f$_{esc}^{H\alpha}$, suggesting that as the escape fraction rises the production efficiency drops. Theoretically speaking, ionising photons are largely produced from young massive stars which spend significant portions of their lives within the birth clouds from which they formed. Escaping photons primarily originate in these systems once escape paths are cleared by feedback mechanisms and by previous ionisation. Standard population models see the production efficiencies of these systems decline rapidly with the ageing population which agrees with our findings \citep{Fire2015}.   

\subsubsection{$\mathrm{M_{UV}}$}
We report significant positive correlations between f$_{esc}^{H\alpha}$ and $\mathrm{M_{UV}}$ at each threshold (Figure \ref{fig:fescMuv}). This is in contrast to the findings from \cite{Papovich2026} and \cite{Giovinazzo2026}, neither of which found correlations in this comparison. Studies such as \cite{Grazian2017} point to fainter sources having larger escape fractions, though with caveats for observational depth, which would agree with our findings. The spectroscopic selection introduces a bias to brighter sources, so it is entirely plausible that this correlation is a product of selection bias. To mitigate potential faint-end selection biases, we further report significant correlations at each threshold with a faint-end limit of $\mathrm{M_{UV}<-19}$ ($\rho$ = 0.34, p$<2\times10^{-2}$ for 0.1\%, $\rho$ = 0.32, p$<8\times10^{-3}$ for 0.01\% and $\rho$ = 0.61, p$<<1\times10^{-6}$ with no threshold). Our results appear to support the faint-galaxy EoR contribution narrative \citep{Kulkarni2019}, as we find increasing f$_{esc}^{H\alpha}$ with fainter sources.

We also contrast our sample to multiple direct observations at varying redshift \citep{Cowie2009,Siana_2010,Steidel2018,Wang_2025} and to simulated estimations \citep{Finkelstein2019} at high redshift as a benchmark for our findings. These direct studies tend to use the ratio of detected photons at 880-912\AA\, normalised by the flux found at the non-ionising UV 1500\AA\, mark to calculate the escape fraction \citep{Steidel2001}, making their methodology fundamentally different from the H$\alpha$-based approach adopted here.

We find a range of observed escape fractions in these samples. \cite{Steidel2018} prioritised star-forming Lyman Break Galaxies and AGN at z$\sim3$, reporting a 12$\pm$0.1\% escape fraction across the whole sample using their Intergalactic + Circumgalactic Medium (IGM+CGM) absorption calculation. \cite{Wang_2025} stacks two sets of 28 galaxies with average redshifts of $\sim$2.4 and $\sim$2.7, finding no significant ($2\sigma$) detection of LyC flux in either stack, and setting $1\sigma$ upper limits of 5\% and 26\% respectively. These observations indicate much larger escape fractions than is determined in most of our sample, possibly due to the sample selection and survey design of ground based observation as noted in \cite{Finkelstein2019}. At lower redshift, \cite{Siana_2010} found a stacked upper limit of 2\% while \cite{Cowie2009} found no significant detection when estimating the relative escape fraction. The relative escape fraction differs from the absolute escape fraction by not including the dust correction factor, so these upper limits represent a generous depiction of LyC escape. Our sample median of 2\% escape fraction agrees more closely with the findings of \cite{Siana_2010} and \cite{Cowie2009}, as well as with the assumed trajectory of the \cite{Finkelstein2019} simulation, found to approach 0.01\% at z$\sim4$. Together, these comparisons suggest that characteristic, population-averaged escape fractions inferred from non-LyC tracers, as well as from non-targeted LyC studies, are typically much lower than those measured in observations designed to identify the strongest LyC-leaking galaxies.

\subsubsection{stellar parameters}
We do not find a significant correlation between the f$_{esc}^{H\alpha}$ and stellar mass besides when considering the full sample (see Table \ref{tab:fesc_threshold_corr} and Figure \ref{fig:fesc_physical}). Theoretically, low stellar masses correlate with high escape fractions due to the shallow gravitational potential wells produced by the halo \citep{Finkelstein2019}. However, \cite{thesan} finds no such clear trend, reporting a largely ambivalent relationship between stellar mass and ionising photon escape. We note that the median stellar mass of the galaxies below the 0.01\% threshold ($\mathrm{log_{10}(M_*/M_\odot = 9.64^{+1.09}_{-1.16})}$) is higher than those above it ($\mathrm{log_{10}(M_*/M_\odot = 9.08^{+0.71}_{-0.48})}$) which extends the contributing mass range and potentially leads to the theoretically expected negative correlation.

We find significant negative correlations of f$_{esc}^{H\alpha}$ to the star formation rate averaged over the last 10Myr ($\mathrm{log_{10}(SFR_{10}/(M_\odot/yr))}$, with increasing magnitude with the inclusion of the full sample. We also find that changing the averaging to 100 Myr significantly diminishes this trend, only becoming significant without thresholds ($\rho$ = -0.56, p$<<1\times10^{-6}$). We report a slight negative correlation of f$_{esc}^{H\alpha}$ to the specific SFR (sSFR/yr) at a threshold of 0.01\% f$_{esc}^{H\alpha}$.  The short timescale SFR is used in \texttt{BEAGLE} to determine the population of massive stars which contribute significantly to the ionising budget. These stars also significantly contribute to the H$\alpha$ emission, which in our model is interpreted as higher rates of LyC absorption. As discussed in \cite{Papovich2026}, galaxies with large ionising budgets at z$>4.5$ are found to have low escape fractions. Comparing to simulations at the EoR \citep{thesan} and observational constraints based on high redshift LyC leakers \citep{Naidu2020}, galaxies with a high SFR (100 Myr average) tend to have larger $\mathrm{f_{esc}}$, where we find no correlation within the same $\mathrm{f_{esc}}$ threshold for this SFR timescale. Our results do corroborate those from \cite{Giovinazzo2026} which report no significant correlation to the SFR density.

\subsubsection{Emission line EW}
We do not find significant correlations between f$_{esc}^{H\alpha}$ and the [OIII]5007 EW using either threshold, but do find a slight positive correlation with the full sample (see \ref{tab:fesc_threshold_corr} and Figure \ref{fig:fesco3ha}). The [OIII]5007 EW is not traditionally used as a proxy for $\mathrm{f_{esc}}$, favouring instead the O$\mathrm{_{32}}$ ratio as a more geometrically considerate method \citep{geometryo32}. It is, however, used in selecting extreme emission line galaxies \citep{Jaiswar2024} which may imply LyC leaker behaviour. \cite{Saxena2022} for example, finds that in their LyC leaker sample at $3.11 < z < 3.53$, the $\mathrm{f_{esc}}$ has a slight positive correlation with the combined [OIII]5007+H$\beta$ EW. The blended emission line and sample selection however limit the utility of this direct comparison. 

We find a significant negative correlation between f$_{esc}^{H\alpha}$ and the H$\alpha$ EW for f$\mathrm{_{esc}}>0.1\%$ which flips without a threshold, implying distinct behaviour in the high f$_{esc}^{H\alpha}$ regime. In the low f$_{esc}^{H\alpha}$ regime, dust becomes a dominating component of this calculation while in the high f$_{esc}^{H\alpha}$ regime the H$\alpha$ EW is likely suppressed by photon leakage. The distinct behaviours are therefore likely to be consequences of the f$_{esc}^{H\alpha}$ model's sensitivity to different dominant sources of attenuation across regimes, transitioning from dust-dominated corrections at low escape fractions to geometry- and density-bounded effects that directly reduce nebular line emission at higher escape fractions.     

\section{Summary and Conclusions}
We report a new sample of 230 galaxies between 1.3$<z<$2.6 with robust H$\alpha$ fluxes from the HUDF-N pointings of the pure-parallel JWST/NIRISS program; OutThere in terms of their ionising photon production and escape. We use this sample to test correlations between physical parameters, nebular emission lines and ionising emission properties drawn at higher redshifts with smaller dynamical ranges. 
\begin{enumerate}

\item We find no significant correlation of $\mathrm{\xi_{ion}^{HII}}$ to the UV slope or sSFR, possibly due to the redder and less star forming galaxies of this sample when contrasted with the high redshift studies. We also do not find that the UV magnitude has any notable correlation with $\mathrm{\xi_{ion}^{HII}}$ as found in high redshift samples, indicating that efficient LyC producers being prevalent in fainter sources may not be true of the general population at z$\sim$2.

\item We do not find the expected anti-correlation of $\mathrm{\xi_{ion}^{HII}}$ to stellar mass. The sample is particularly scattered at the high mass end which may prevents a significant correlation being defined. 

\item We corroborate the findings of nebular emission line comparisons to $\mathrm{\xi_{ion}^{HII}}$ both at the same and higher redshifts and extend the dynamical range of equivalent widths to the lowest range. We find both in the case of [OIII]5007 and H$\alpha$ that an increasing EW correlates with an increasing $\mathrm{\xi_{ion}^{HII}}$.

\item We find no correlations between the UV slope derived escape fraction and redshift and find similar values to those reported at z$\sim$2 when using the \cite{Chisholm2022} UV slope correlation. We find that our theoretical $\mathrm{f_{esc}}$'s are more varied, finding both much larger and much smaller escape fractions than those found by the empirical relation though still within theoretical limits. We do find tentative correlations to redshift in these methods, however, these methods are purely theoretical and require further testing. 

\item We find consistent positive correlations between the H$\alpha$ $\mathrm{f_{esc}}$ and M$\mathrm{_{UV}}$, indicating that fainter galaxies may have more notable LyC escape fractions.

\item We find that our escape fractions do not correlate with stellar mass or [OIII]5007 \AA\,EW above the 0.01\% escape threshold. Intuitively the lower mass and high stellar activity indicators suggest shallow gravitational wells and hard ionising emissions potentially forming escape routes. However, our $\mathrm{f_{esc}}$ limits remove the highest mass and lowest [OIII]5007 emitters which contribute to the correlation. 

\item The H$\alpha$ $\mathrm{f_{esc}}$ correlation is threshold dependent, with the direction reversing for the lowest $\mathrm{f_{esc}}$ galaxies, indicating more complex dependencies.

\item The short term SFR and the model derived $\mathrm{\xi_{ion}^{BEAGLE}}$ do consistently anti-correlate with the escape fraction. Both are effective proxies of the massive stellar population and are strongly interdependent \citep{Jaiswar2026a}, and thus depict similar behaviour. The anti-correlation suggests the escape fraction favours systems with established stellar populations rather than young ones. 

\item The $\mathrm{\xi_{ion}^{HII}}$ does not correlate with the escape fraction at any threshold. Our $\mathrm{f_{esc}}$ is strongly dependent on and limited by dust, while the $\mathrm{\xi_{ion}^{HII}}$ is an intrinsic parameter once dust corrected. These results suggest that the conditions producing this radiation and the conditions for its escape are largely independent for this sample. 

\end{enumerate}
This work demonstrates the value of a large, uniformly selected intermediate-redshift sample as a benchmark for ionising photon production and escape. By showing how key trends in physical properties and nebular line strengths extend to lower redshifts and wider dynamical ranges, we provide critical validation of relations extrapolated to the EoR. We emphasise the diversity of ionising efficiencies across galaxy populations and the scatter at the extremes. This work also warns against the use of individual predictors of ionising emission, as we note their dependence on the methodology used in deriving both the data and the result. 

\section{Acknowledgements}
This research was partly supported by the Australian Research Council Centre of Excellence for All Sky Astrophysics in 3 Dimensions (ASTRO 3D), through project number CE170100013. CMT was supported by an ARC Future Fellowship under grant FT180100321.
\\The International Centre for Radio Astronomy Research (ICRAR) is a Joint Venture of Curtin University and The University of Western Australia, funded by the Western Australian State government.
\\D.W and Z.H. acknowledge support for program JWST-GO-03383, provided by NASA through a grant from the Space Telescope Space Institute, which is operated by the Associations of Universities for Research in Astronomy, Incorporated, under NASA contract NAS5-26555

\begin{appendix}
\appendix
\renewcommand{\thefigure}{A\arabic{figure}}
\setcounter{figure}{0}
\renewcommand{\thetable}{A\arabic{table}}
\setcounter{table}{0}
\section{}\label{AppendixA}
\begin{figure}
    \centering
    \includegraphics[width=1\linewidth]{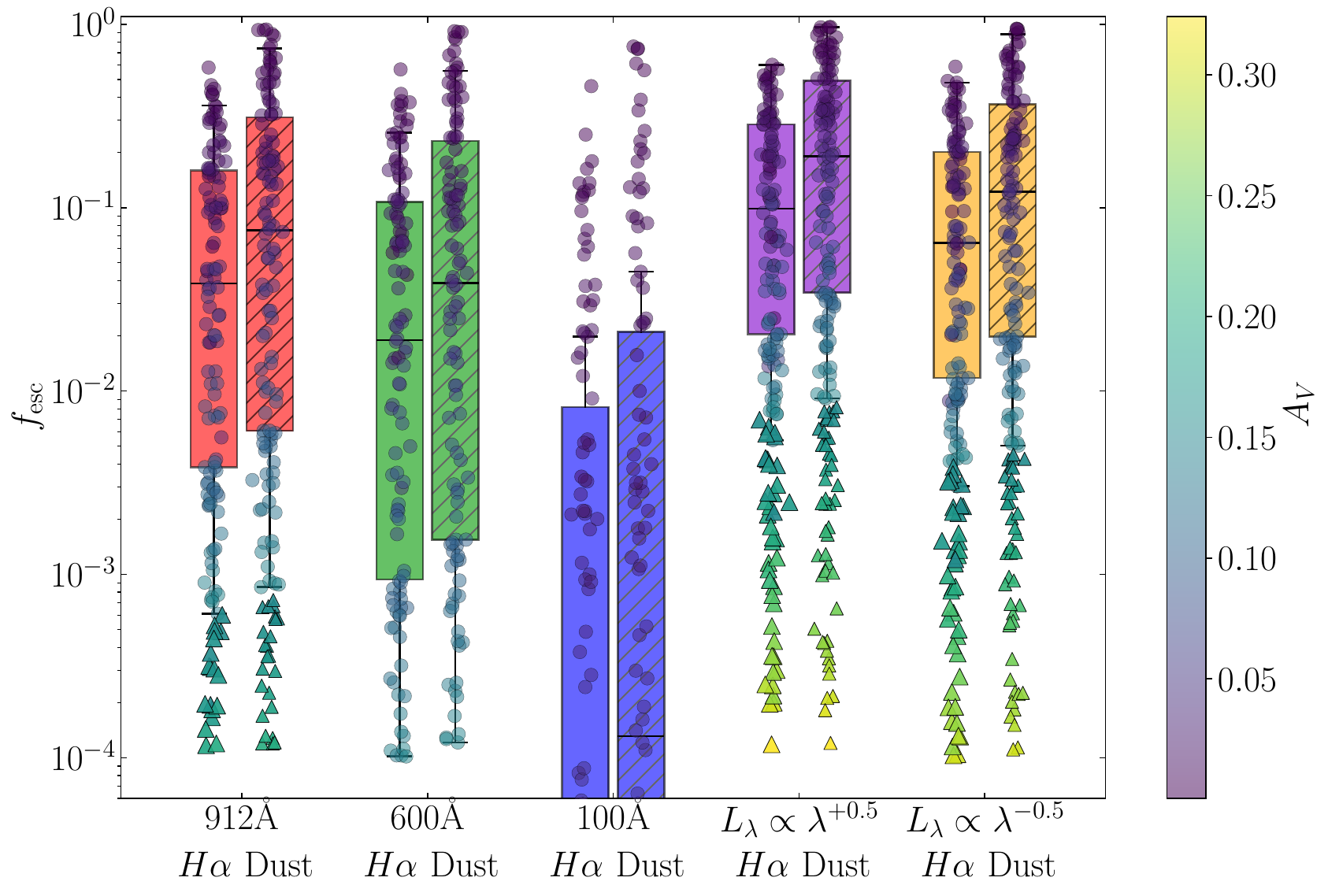}
    \caption{The LyC dust attenuation approximation comparison. From left to right, the panels show the derived $f_{\mathrm{esc}}^{H\alpha}$ (plain) and $f_{\mathrm{esc}}^{\mathrm{dust}}$ (crosshatched) escape fractions for different assumptions about the intrinsic LyC spectrum. The distributions are calculated using a common sample selected by the $L_\lambda$ flat spectrum evaluated at 600\AA\, with $f_{\mathrm{esc}}>10^{-4}$ (0.01\%), ensuring a consistent population comparison between attenuation prescriptions. The first three cases assume a flat $L_\lambda$ spectrum evaluated at 912, 600 and 100\AA, while the final two adopt power-law spectra with $L_\lambda\propto\lambda^{+0.5}$ and $L_\lambda\propto\lambda^{-0.5}$, respectively. Boxes show standard Tukey statistics (25th-75th percentile range, median line, and 1.5$\times$IQR whiskers), while individual galaxies are shown as circles coloured by their $A_V$. Galaxies excluded from the common sample that re-enter the detectable regime ($f_{\mathrm{esc}}>10^{-4}$) under alternative LyC spectral assumptions are shown as triangles. The corresponding medians and percentile ranges are reported in Table~\ref{tab:lyc_slope_comparison}. 
    }
    \label{fig:appendix_dust}
\end{figure}

Here we discuss the choice of a flat L$_\lambda$ with a reference wavelength of 600\AA\, in the calculation of f$\mathrm{_{esc}^{dust}}$ and briefly explore alternatives to this choice. We refit the sample using flat spectra referenced at either integration limit (912\AA\, and 100\AA\,) as well as considered evolving L$_\lambda$ (see Figure \ref{fig:appendix_dust} and Table \ref{tab:fesc_threshold_corr}). Of the flat spectra, a choice of 912\AA\, as the reference point results in less attenuation overall and thus larger escape fractions. \cite{Chisholm2018} discusses the significance of dust attenuation at 912\AA\, in reference to an idealised model similar to this one, and it is a frequent generous choice to use the Lyman limit as a reference point for LyC escape calculations \citep{Steidel2001,Cooke2014}. \cite{Siana_2007} notes the ionising emissions of young stars in stellar population synthesis modelling \citep{LindaSmith} do not significantly evolve between the commonly referenced 900\AA\, range and their chosen 700\AA\, where the most significant drop-off is below 500\AA . We find that using a 912\AA\, reference point increases the median calculated escape fraction f$\mathrm{_{esc}^{H\alpha}}$, from 1.9\% to 3.9\% and has a wider overall range of values. We also find that choosing the lower integration limit of 100\AA\, brings the median f$\mathrm{_{esc}^{H\alpha}}$ only slightly above the chosen threshold, as the attenuation by this wavelength is notably strong.

\begin{table}
\centering
\caption{Median$\pm$IQR escape fractions obtained under different assumptions for the intrinsic LyC spectrum using the common sample described in Figure \ref{fig:appendix_dust}. The $f_{\rm esc}^{\mathrm{H}\alpha}$ and $f_{\rm esc}^{\rm dust}$ values represent the medians and IQRs of the distributions shown in Figure \ref{fig:appendix_dust}.}
\label{tab:lyc_slope_comparison}
\begin{tabular}{lccc}
\hline
LyC reference & Spectral shape & Median $f_{\rm esc}^{\mathrm{H}\alpha}$\% & Median $f_{\rm esc}^{\rm dust}$\% \\
\hline
912\,\AA{}  & $L_\lambda = \mathrm{constant}$ & $3.9^{+12.1}_{-3.5}$ & $7.5^{+23.5}_{-6.9}$ \\
600\,\AA{}  & $L_\lambda = \mathrm{constant}$ & $1.9^{+8.9}_{-1.8}$ & $3.9^{+19.2}_{-3.7}$ \\
100\,\AA{}  & $L_\lambda = \mathrm{constant}$ & $0.01^{+0.81}_{-0.01}$ & $0.01^{+2.09}_{-0.01}$ \\
integrated slope & $L_\lambda \propto \lambda^{+0.5}$ & $9.9^{+18.5}_{-7.8}$ & $19.0^{+30.2}_{-15.6}$ \\
integrated slope & $L_\lambda \propto \lambda^{-0.5}$ & $6.4^{+13.7}_{-5.2}$ & $12.2^{+24.5}_{-10.2}$ \\
\hline
\end{tabular}
\end{table}

Integrating the LyC attenuation over a finite wavelength range requires an assumption about the intrinsic spectral shape below the Lyman limit. Power-law approximations are commonly adopted in ionising background models, with spectra such as L$_\nu \propto \nu^{-1.5}$ used for AGN and quasar continua \citep{Madau96}, corresponding to L$_\lambda \propto \lambda^{-0.5}$. We additionally test both this commonly adopted slope and a redder L$_\lambda \propto \lambda^{+0.5}$ prescription. In both cases, the resulting median f$\mathrm{_{esc}^{H\alpha}}$ values exceed those derived using the flat 912\AA, approximation. Since the flat spectrum normalised at 912\AA, provides the least attenuated possible estimate within our wavelength range, these steeper spectral assumptions produce escape fractions beyond the expected upper limit of this approximation. We therefore consider them unlikely to provide a physically representative description of the intrinsic LyC spectrum for our sample. Although the intrinsic LyC spectrum is expected to depend on the underlying stellar population properties, we find that the flat spectral assumption introduces only limited systematic variation in the derived escape fractions compared with alternative power-law prescriptions. A full treatment would require coupling the attenuation calculation with a stellar population synthesis model.

\end{appendix}

\FloatBarrier

\bibliography{sample631}{}
\bibliographystyle{plainnat}

\end{document}